\documentclass{aa}

\usepackage{graphicx}
\usepackage{txfonts}
\usepackage{amssymb}
\usepackage{listings}
\usepackage{nameref}
\usepackage{rotating}
\usepackage[utf8]{inputenc}
\usepackage[draft]{hyperref}
\usepackage{booktabs}
\usepackage{color}
\usepackage{adjustbox}

\newcommand{\rev}[1]{{\color{black} #1}}

\begin{document}

    \title{
    Drifting Features: Detection and evaluation in the context of automatic RRLs identification in \textsc{VVV}
    }
    \subtitle{}

   \author{J. B. Cabral\inst{1, 2}
          \and
          M. Lares\inst{2}
          \and
          S. Gurovich\inst{2}
          \and
          D. Minniti\inst{3,4,5}
          \and
          P.M. Granitto\inst{1}
          }

   \institute{
        Centro Internacional Franco Argentino de Ciencias de la
        Informaci\'on y de Sistemas (CIFASIS, CONICET--UNR) \\
        \email{cabral@cifasis-conicet.gov.ar}
        \and
        Instituto De Astronom\'ia Te\'orica y Experimental -
        Observatorio Astron\'omico C\'ordoba (IATE--OAC--UNC--CONICET)
        \and
        Departamento de Física, Facultad de Ciencias Exactas, Universidad Andrés Bello, Av. Fernandez Concha 700, Las Condes, Santiago, Chile
        \and
        Instituto Milenio de Astrofísica, Santiago, 7500912, Chile
        \and
        Vatican Observatory, V-00120 Vatican City State, Italy       
    }

   \date{}

\abstract
{As most of the modern astronomical sky surveys produce data faster than humans can analyze it, Machine Learning (ML) has become a central tool in Astronomy. Modern ML methods can be characterized as highly resistant to some experimental errors. However, small changes on the data over long \rev{angular distances} or long periods of time, which cannot be easily detected by statistical methods, can be harmful to these methods.}
{We develop a new strategy to cope with this problem, also using ML methods in an innovative way, to identify these potentially harmful features.}
{We introduce and discuss the notion of Drifting Features, related with small changes in the properties as measured in the data features. We use the identification of RRLs in VVV based on an earlier work and introduce a method for detecting Drifting Features. \rev{For the VVV, each sky observation zone is called a tile.} Our method forces the classifier to learn from the sources (mostly stellar 'point sources') the tile they are originated, and select the features more relevant to the task of finding candidates to Drifting Features.}
% Our method forces a classifier to learn the tile of origin of diverse sources (mostly stellar 'point sources'), and select the features more relevant to the task of finding candidates to Drifting Features.} # old
%
{We show that this method can efficiently identify a reduced set of features that contains useful information about the tile of origin of the sources. For our particular example of detecting RRLs in VVV, we find that Drifting Features are mostly related to color indices. On the other hand, we show that, even if we have a clear set of Drifting Features in our problem, they are mostly insensitive to the identification of RRLs.}
{Drifting Features can be efficiently identified using ML methods. However, in our example, removing Drifting Features does not improve the identification of RRLs.}

% 5 {} token are mandatory

        
  \maketitle

% ================================================================ ==================
% INTRO
% ==================================================================================

\section{Introduction}

Most of the modern astronomical sky surveys are characterized by fast pace data ingestion, data intensive science cases or automatic reduction pipelines \citep[e.g.][]{feigelson2012big}, which often lie on the verge of technological developments and analysis capabilities.
This unprecedented availability of observations challenges the traditional approaches for data analysis, leading to a shift in the paradigm for knowledge discovery \citep{bell2009beyond}, notably dominated by machine learning (ML) techniques \citep{ball2010data}.
Albeit the difficult mathematical and statistical foundations, a complex terminology driven by the confluence of several sciences and the arduous interpretation of the results, the training of intelligent agents has become an every-day practice in astronomy. The accessibility of easy-to-use free software resources mostly written for \textit{R} \citep{team2000r} or \textit{Python} \citep{van2003python} languages was fundamental to this step. 

In most cases ML methods can be separated into two basic steps: first, raw data is converted into a set of useful features that are relevant to the task at hand (e.g. periods or intensities) and then these features are fed to a classifier or a statistical method \citep[see e.g. ][]{mitchell1997machine}.

ML methods have a number of limitations. For instance, they are highly susceptible to errors produced by the limitations of the datasets \citep{cai2015challenges}.
The results can also be hampered by the not fully understood role of the features \citep{duboue2020art} or by  the biases introduced by improperly defined experiments \citep{domingos2012few}.
These facts are well known and have not been ignored in the astronomical community \citep{2020MNRAS.492.5377L}.

Here we are interested in the role of some sources of noise that are present in commonly used features in astronomical research, and their impact on the results of ML methods in this context. We use data from the synoptic survey ``\textit{Vista Variables in Via L\'{a}ctea}'' (VVV, \citealp{2010NewA...15..433M}), observed with the Vista telescope \citep{sutherland2015visible}, which pursues, among its main objectives, to produce a three-dimensional map of a large part of the galactic center (Bulge) of the Milky Way and a fraction of the internal Galactic Disk. The VVV data is presented in units called ``tiles'', which are rectangular areas of the sky surveyed over time. For each tile, the VVV data reduction pipeline \citep{emerson_vista_2004} provides a pre-processed image and a database of files with the values of position, magnitude and color indices of the light sources present in the image, which comprises the ``photometric catalog''. These catalogs are the main subject of this study.

The images are subject to two noise sources, namely, experimental errors and observation conditions. The derived catalogs are also affected, since the noise permeates all the survey information which is comprised on a set of features or observables. Atmospheric conditions, moon phases, maintenance of the camera and telescope or modifications to the software, among many others, can influence the observation or recording of the data. As a consequence, the derived measurements that are used as features for a ML analysis can be also prone, at different levels, to these errors and conditions. 

Random measurement errors are present in every experimental or observational science
They are unavoidable, but each error typically affects only a single or a reduced set of observations\rev{, in particular wide-field surveys images can be affected by issues such as weather, astronomical conditions, software updates, etc.} ML methods can cope efficiently with this kind of errors. For a large survey as VVV, observational conditions can change slightly (but not randomly) over long periods of time or for different regions in the sky.
This problem is more difficult for ML methods.
In many situations we want to train an intelligent agent using a well--known portion of the survey, and then use it to predict other less--known zones searching for a given astronomical phenomenon.
Given the ML methodology, the agent will work efficiently on training data, but will probably fail to generalize to other zones due to this slight change in observational conditions.
Due to the diverse nature of the features extracted from the data (intensity, periods, colours, etc.), possibly they will reflect this effect in different proportions. 
It is then interesting to ask whether it is possible to automatically detect which of the extracted features are more sensitive to these changes in observational conditions.
Hereafter we call ``Drifting Features'' to the ones in a dataset that are sensitive to the observational conditions. We aim at evaluating their influence over a large scale ML experiment.

As a working example, we focus on the problem of detecting RR-Lyrae (RRLs) variable objects over VVV data. That is, we train classifiers using data from some VVV tiles, and evaluate them in the task of identifying RRLs on other tiles.

Drifting Features should be consistent within a limited zone of the sky (for instance, a tile or two consecutive tiles) but should show slight changes, almost undetectable by most simple statistics, between tiles that are separated apart\footnote{\rev{Whether two consecutive tiles or, in general, two regions on the sky are similar depends on the survey strategy. For example, regions that are repeatedly observed in quick succession in a survey, or that share important observational parameters such as the exposure time, could show this property even if they have long angular distances. The opposite is also valid, regions that are close at angular distances can be observed under very diverse conditions.}}. Those changes could potentially alter the capabilities of the classifier. To detect these features and their effect on automatic classification, we propose to use, again, ML methods. If we face a ML method with the task of discriminating data from two tiles, it will be forced to learn differences among the tiles that are present in the features. We can then use Feature Selection methods \citep{guyon2002gene} to evaluate the importance of each feature for this classifier that discriminates tiles, and mark highly relevant features as candidates of been Drifting Features. \rev{In other words, we propose to learn a separate task (the tile of origin of a source) not because it is unknown or difficult, but as a method to detect which are the features that are more useful to this task, the features that contain information that change with the tile of origin.}

This work is divided into the following sections: in Section~\ref{section:vvv} 
we explain our experimental setup (data, feature extraction, model selection,etc.). In Section \ref{section:drift} we introduce our procedure for the identification of Drifting Features and in Section \ref{section:evaluation} we evaluate the effect of these features in the task of RRL identification. Finally in Section \ref{section:conclusions} we discuss our results and draw our conclusions.

% =============================================================================
% SAMPLING
% =============================================================================

\section{Experimental setup}
\label{section:vvv}

\subsection{Data}

One of the main objectives of the VVV is the creation of a 3D map of the bulge and the galactic center \citep{2010NewA...15..433M} for which the search for variable stars in general, and RRLs in particular, is important due to their use as standard candles \citep{bailey1902discussion}.
To this end, the survey relies on data from the VIRCAM infrared camera, mounted on ESO's VISTA survey telescope \citep{sutherland2015visible}, which at the time of its construction was the largest NIR camera with 16 non-contiguous \mbox{2k x 2k} detectors.
To complete a contiguous tile, VIRCAM simultaneously exposes its detectors 6 times with a suitable offset. Each one of the exposures is called a \textit{pawprint} and the combination of the six overlapping pawprints is a \textit{tile}. For this reason each pixel is observed in at least 2 pawprints and also the edges are shared with the observations of the adjacent tiles.
The survey observation plan was organized in two stages: during the first year the tile is observed in five astronomical filters Z, Y, J, H and Ks separated by a few hours; then, in subsequent years, it is re-observed using the Ks band for variability studies. Only some tiles are observed in multi-band after the first year.

The dataset used in this work is the one presented in 
\citet{cabral2020automatic}, which consists of {62} features extracted with \texttt{feets} \citep{cabral_fats_2018}
from light curves that were reconstructed from the photometric catalogs provided by the Cambridge Astronomical Survey Unit (CASU).

\begin{figure*}
\centering
    \includegraphics[width=.7\textwidth]{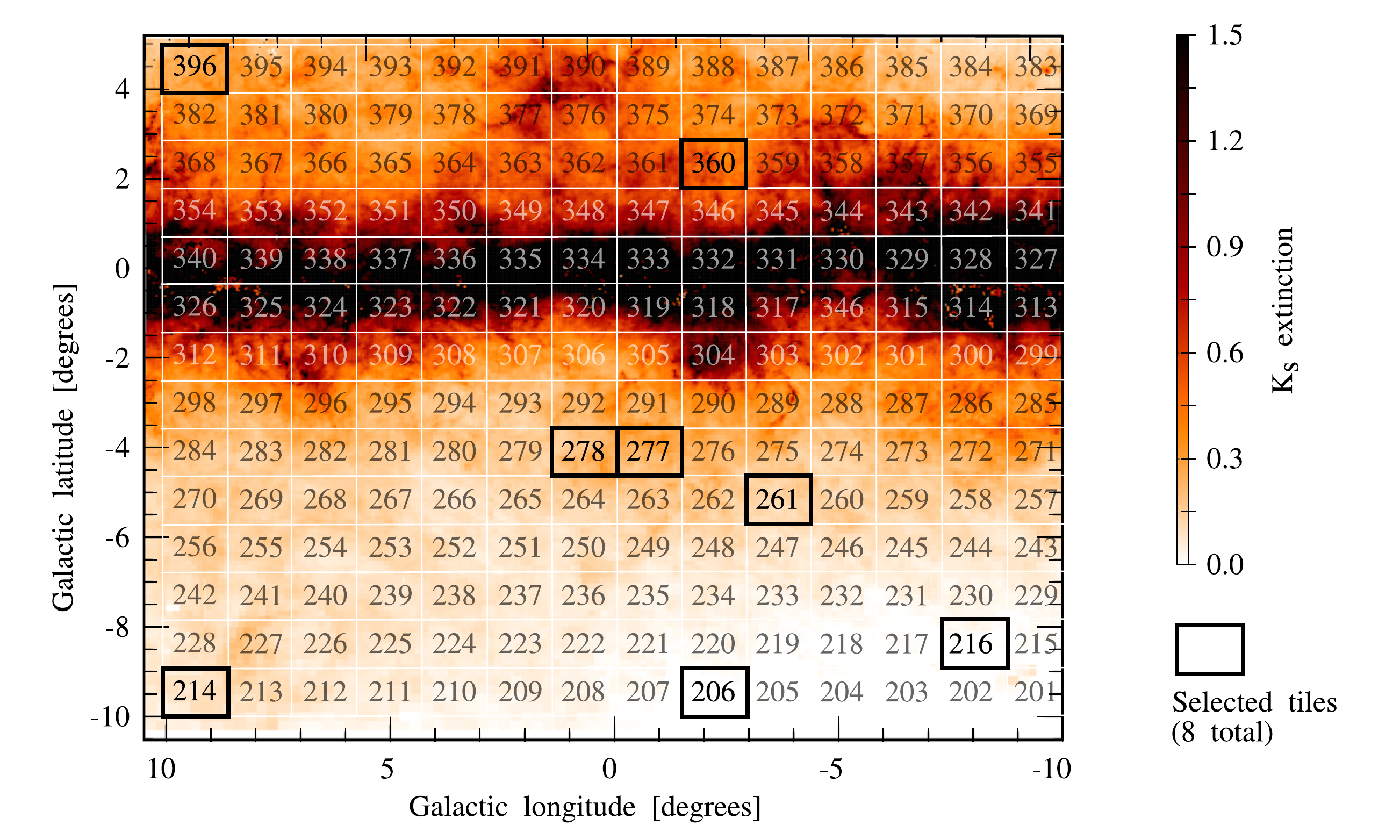}
    \caption{\label{fig:used_tiles}
    Map of the Bulge tiles of the  \textsc{VVV} survey over an extinction map (Extinction map adapted from and \citet{gonzalez2012reddening}). We highlight the tiles used in this work with red borders.}
\end{figure*}

From the original dataset we selected 8 tiles located at different zones of the Bulge, as shown in Figure \ref{fig:used_tiles}. For each tile we extracted all the RRLs plus a uniform sample of $2000$ unclassified, unsaturated and non-faint sources (average magnitude between $12$ and $16.5$).
From these selections, sources with invalid values were removed, leaving the final dataset for this work, described in  Table~\ref{tab:sample}.

We choose to use a reduced dataset with around $2000$ sources for each tile in order to decrease dramatically the computational burden of our experiments. As shown in \citet{cabral2020automatic}, the use of a reduced dataset can lead to optimistic estimations of the accuracy of the detections, but our main objective is to find and characterize the features that best represent the differences between the tiles and not the accuracy of the detection of the RRLs.

\begin{table}
\caption{Total number of sources, RRL and sample taken in each tile used in this work.
}
    \centering
\begin{tabular}{l|rrr}
\toprule
 \textbf{Tile} &    \textbf{Total} &  \textbf{RRL} &  \textbf{Sample} \\
%\midrule
 b206 &   407720 &   47 &    2047 \\
 b214 &   376822 &   35 &    2034 \\
 b216 &   334773 &   43 &    2043 \\
 b261 &   735838 &  253 &    2252 \\
 b277 &   831323 &  430 &    2429 \\
 b278 &   857887 &  437 &    2436 \\
 b360 &  1029149 &  679 &    2669 \\
 b396 &   729671 &   15 &    2015 \\
\bottomrule
\end{tabular}
\label{tab:sample}
\end{table}

% =============================================================================
% MODEL SELECTION
% =============================================================================
\subsection{Error measures}

We face two different binary classification problems in this work.
First we try to separate sources between two tiles, this instance is made in order to construct a Tile Classifier (hereafter TC) that allows to asses the relevance of the features. \rev{As we stated in the Introduction, we use the TC as an auxiliary method that allow us to detect which features are candidates to Drifting Features.}
Then, we build a Source Classifier (hereafter SC) that seeks to discriminate RRL sources from unknown sources.
In the first problem both classes are nearly balanced in all cases. On the other hand, as discussed in \citet{cabral2020automatic}, the identification of a few variable stars within a large set of unknown sources is usually a highly imbalanced problem, which generates several inconveniences such as those discussed in the recent work by \citet{hosenie2020imbalance} and requires specific error measures.

In the RRL detection problem (SC) we will call RRL samples as the positive class and the other sources as the negative class.
In the tile identification problem (TC) both classes (the two tiles) are equivalent, so we will arbitrarily call one of them as positive and the other as negative.
All positive samples (in this case, either a source or a tile) that are correctly identified by the classifier are called true positives (TP), otherwise if they are missed by the classifier they are called false negatives (FN).
Negative samples that are wrongly classified are called false positives (FP), and those correctly identified are called true negatives (TN).
Using a combination of these four outcomes, we can define two complementary performance measures, called Precision and Recall, which are adequate to deal with unbalanced problems. The Precision is defined as $\displaystyle{TP / (TP + FP)}$. It measures, for example, the fraction of real RRLs detected over all those retrieved by the classifier. The Recall, on the other hand, is defined as $\displaystyle{TP / (TP + FN)}$. It measures, in the same example, the fraction of all RRLs that are detected by the classifier.

Many classifiers can change their decisions outputs by adjusting the probability threshold that considers an observation to be positive or negative.
A high threshold increases the Precision and decreases the Recall since fewer cases are classified as positive, while a low threshold generates the opposite effect.
To evaluate Precision and Recall together, we consider the Precision-Recall curves, where we plot a set of pairs of values corresponding to different thresholds.
A curve that approaches the top-right corner is, in general, considered to represent a better classifier.

For balanced classification problems it is common to find more traditional metrics in the literature. According to that, for the tile identification problem we also use Accuracy $\displaystyle{(TP + TN) / (TP + FP + TN + FN)}$ and the area under the ROC curve (ROC-AUC) measures. The ROC curve is equivalent in concept to the Precision-Recall curve described before, and the area under it is a global measure of the performance of the classifier. The only difference between both curves is that a ROC curve that approaches the top-left corner represents a better classifier.

\subsection{Model Selection}
\label{section:model_select}

For the TC problem we evaluated four classifiers with diverse foundations: SVM (Support Vector Machine) with linear kernel \citep{vapnik2013nature}, SVM with Radial Basis Function (RBF) kernel, K-Nearest Neighbors \citep[KNN, ][]{mitchell1997machine} and Random Forest \citep[RF, ][]{breiman2001random}, \rev{all implementations from the Scikit-Learn Python package \citep{pedregosa2011scikit}}.

To determine the best hyper-parameters for every model, we executed a grid-search of all possible combinations of values for each hyper-parameter over a fixed list. We used a \rev{5 k-fold} setup on a dataset with tiles b278 and b261, using precision as performance measure. These tiles were chosen  because they are not extreme in location or in their balance, such as b396 or b220. 

With this setup, we selected the following hyper-parameter values:
\begin{description}
    \item[SVM-Linear:] $C = 100$.
    \item[SVM-RBF:] $C = 100$ and $\gamma = 0.003$.
    \item[KNN:] $K = 56$ with a $Manhattan$ metric; also, the importance of the neighbor class was not weighted by distance.
    \item[RF:] We created $500$ decision trees with Information-Gain as metric, the maximum number of random selected features for each tree is the $0.5$ of the total number of features, and the minimum number of observations in each leaf is $5$.
\end{description}

\begin{table}
    \caption{
    Classification metrics of the SVM models (with kernel linear and RBF), RF and KNN  on the tiles b278 and b261 sources.}
    \centering
\begin{tabular}{l|rrr}
\toprule
      \textbf{Model} &  \textbf{Precision} &  \textbf{Recall} &    \textbf{AUC} \\
\midrule
 SVM-Linear &     0.8511 &  0.8511 & 0.9286 \\
    SVM-RBF &     0.8003 &  0.8003 & 0.8680 \\
         RF &     0.7707 &  0.7707 & 0.8548 \\
        KNN &     0.6973 &  0.6973 & 0.7685 \\
\bottomrule
\end{tabular}
    
\label{tab:model_select}
\end{table}

\begin{figure*}
    \centering
    \includegraphics[width=0.8\textwidth]{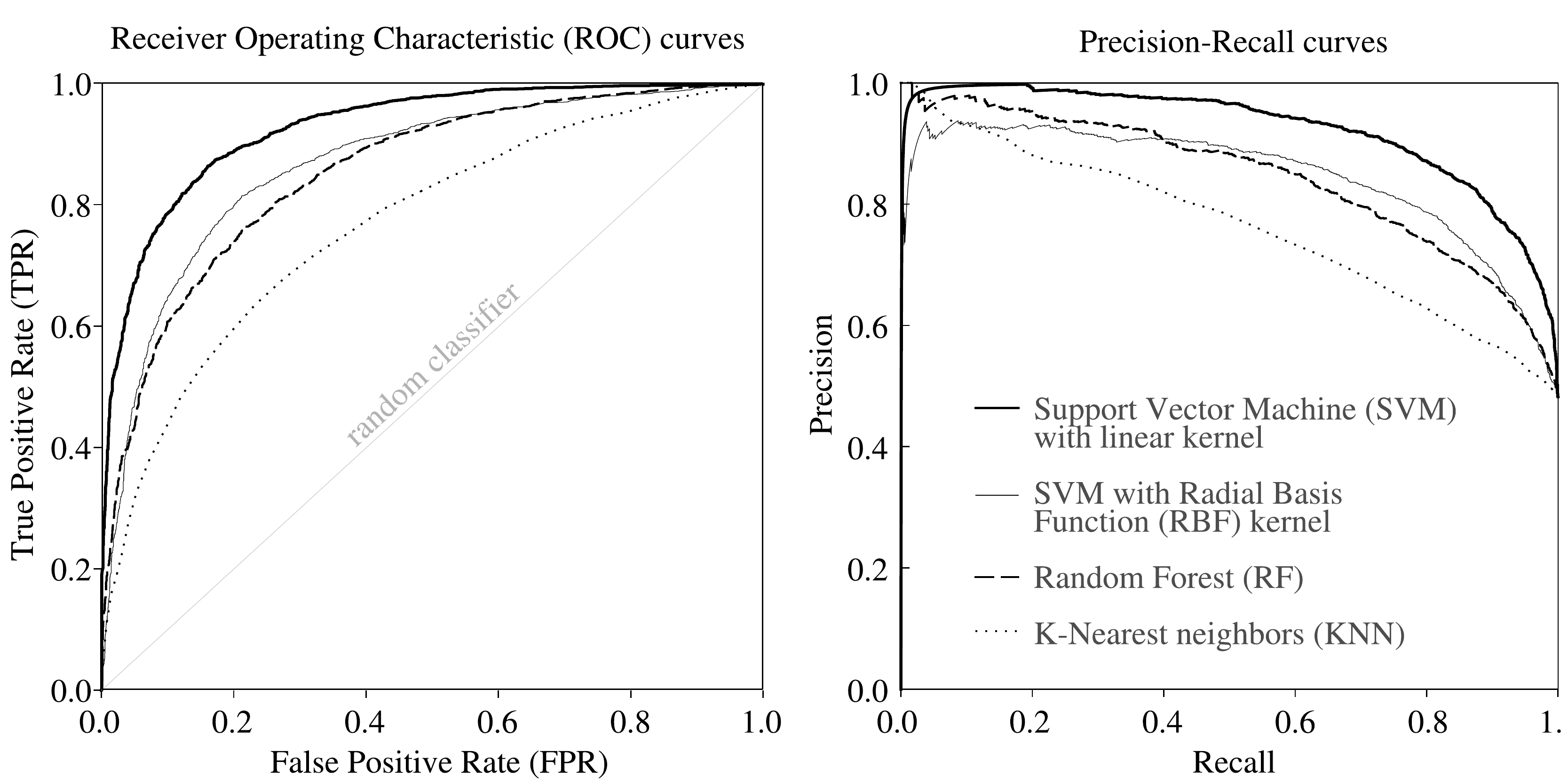}
    \caption{\textsc{ROC} (left) and Precision-Recall (right) curves of the  SVM models (with linear and RBF kernels), RF and KNN, for the prediction of the tile of a given source, using 10--fold CV with tiles b278 and b261.}
    \label{fig:model_select}
\end{figure*}

Using the optimal values for the hyper-parameters we compared the four models on the same dataset using a \rev{10--fold cross--validation} setup. Table~\ref{tab:model_select} shows the corresponding results using the default threshold (0.5) for all models. For all three metrics considered (Precision, Recall and AUC) the SVM-Linear classifier clearly outperformed all the other classifiers. More importantly, Figure \ref{fig:model_select} shows the corresponding ROC and Precision-Recall curves, which show that SVM-Linear also outperforms the other methods for all possible thresholds.
Given these results, we selected SVM-Linear as the classifier for the tile identification problem.

For the SC problem \citet{cabral2020automatic} already determined that RF is the classifier with the best performance for our dataset and general experimental setup. 
% =============================================================================
% Feature selection
% =============================================================================
\subsection{Feature Selection}
\label{subsection:rfe}

% Inspired by: https://arxiv.org/pdf/2001.04081.pdf

Feature selection \citep{guyon2013introduction} is the process of extracting some subsets of features from the entire set in order to optimize the classification performance and/or the computational complexity of the problem. 
We choose for this work the Recursive Feature Elimination algorithm \citep[RFE, ][]{guyon2002gene}. The method is widely adopted, characterized by its good performance and simplicity. As a backward selection method, RFE starts with all the features and sequentially eliminates the unimportant features using a recursive process. 

RFE is integrated with a classification method, which provides at each step of the recursion the importance score of the features. RFE iteratively executes the underlying classifier and extracts the score for each variable, then the variable (or group of variables) with worse performance (according to the score) is eliminated. 

The method typically ends when the desired (fixed) number of features to select is reached. Another possibility is to monitor a performance metric for the subsets (for example the accuracy on an independent validation set) and stop the recursion when the metric is optimal.

% paragraph based on: https://github.com/scikit-learn/scikit-learn/blob/95119c13a/sklearn/feature_selection/_rfe.py#L587-L589
In this work we rely on the RFE implementation with k-fold Cross-Validation (RFECV) as stopping criteria \rev{also} provided by the Scikit-Learn package \citep{pedregosa2011scikit}. 
RFECV produces $k$ replicated experiments ($k=5$ in our work), each one selecting features over ($k-1$) folds and monitoring the classification error ($1-accuracy$) over the remaining fold. Then it determines the number of features to select looking for the least average error throughout all the folds. In the last step, RFECV produces a final selection using all the dataset to select the features and stopping at the previously selected point.

It is worth mentioning that the classifier embedded in the RFE in the feature selection stage may be different from the eventual method in the final classification stage.

\section{Finding Drifting Features}
\label{section:drift}

As we stated in the Introduction, we propose to use ML methods to detect Drifting Features, looking for features that are useful to know the tile of origin of a given source (exclusively from features derived from the pawprint stack photometry, without any other Header Keyword Data). With this goal, in this first experiment we consider all the sources in each tile together (RRLs \& unknowns) and train classifiers to learn the tile of origin of each source and not its astronomic type. 

We apply the RFECV method, as described in the previous section, on 28 binary classification problems, consisting each one in separating a different pair of tiles from the set of 8 tiles in our dataset. Thus, for each SC problem (for example, separating tiles b206 and b214) we obtain from RFECV a subset of selected features for that problem. Each subset has potentially a different length, as discussed before. 

\begin{figure}
    \centering
    \includegraphics[width=.85\columnwidth]{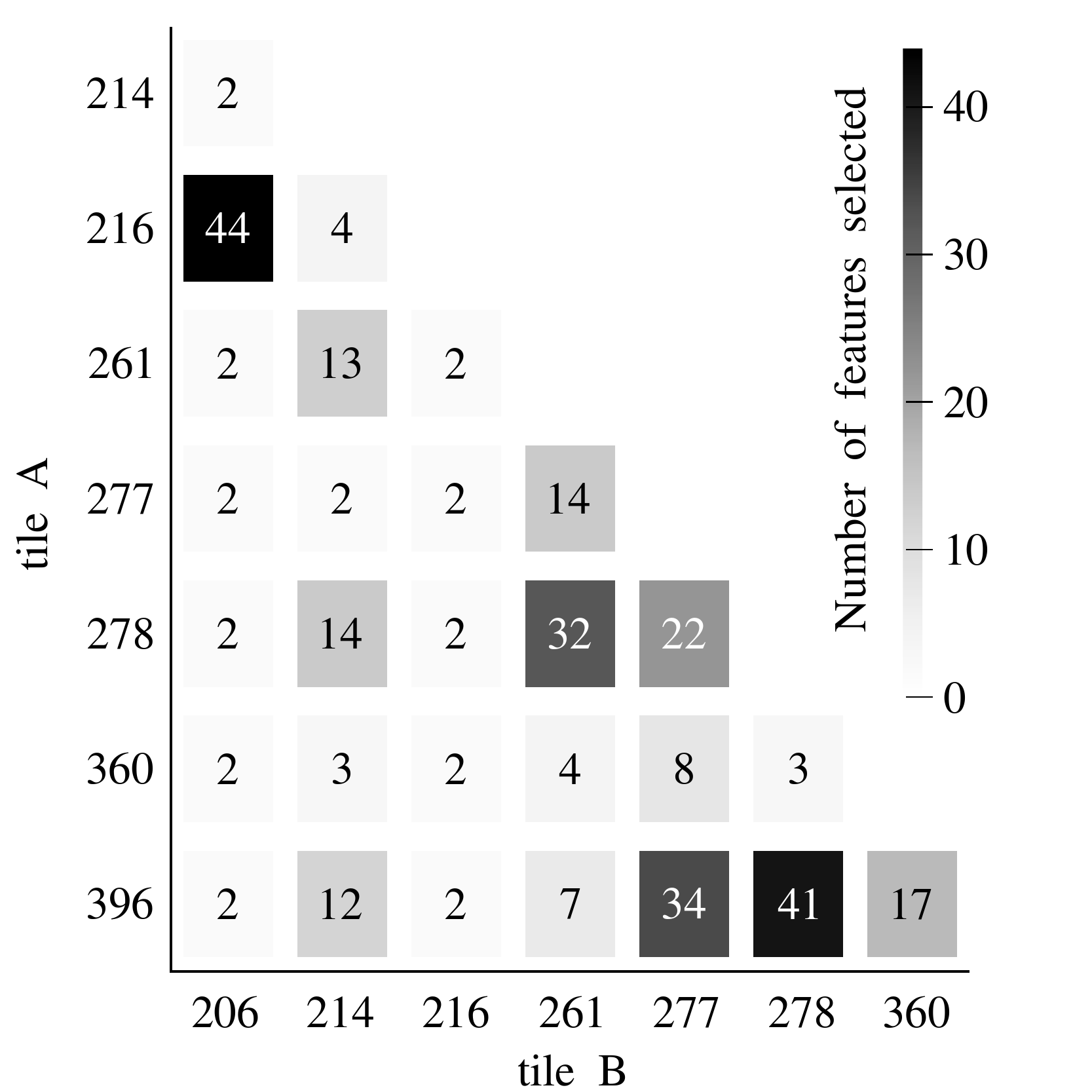}
    \caption{
    Number of features selected by RFE for each binary TC problem. Each cell corresponds to the dataset including Tile A (Rows) and Tile B (Columns).}
    \label{fig:lless_heatmap}
\end{figure}

Figure \ref{fig:lless_heatmap} shows the number of features selected for each problem. It is evident that there are two different behaviours. In some cases RFE selects just a few features, as for example tile b216 with any other except b206; opposite, in some other cases the selected subset contains a high number of features (tile b206 against b216, b396 against b277 or b278, for example). But the number of selected features by itself is not relevant, what is more important to identify Drifting Features is how well they separate the two tiles. 

\begin{figure*}[tbh!]
    \centering
    \includegraphics[page=1, width=.7\textwidth]{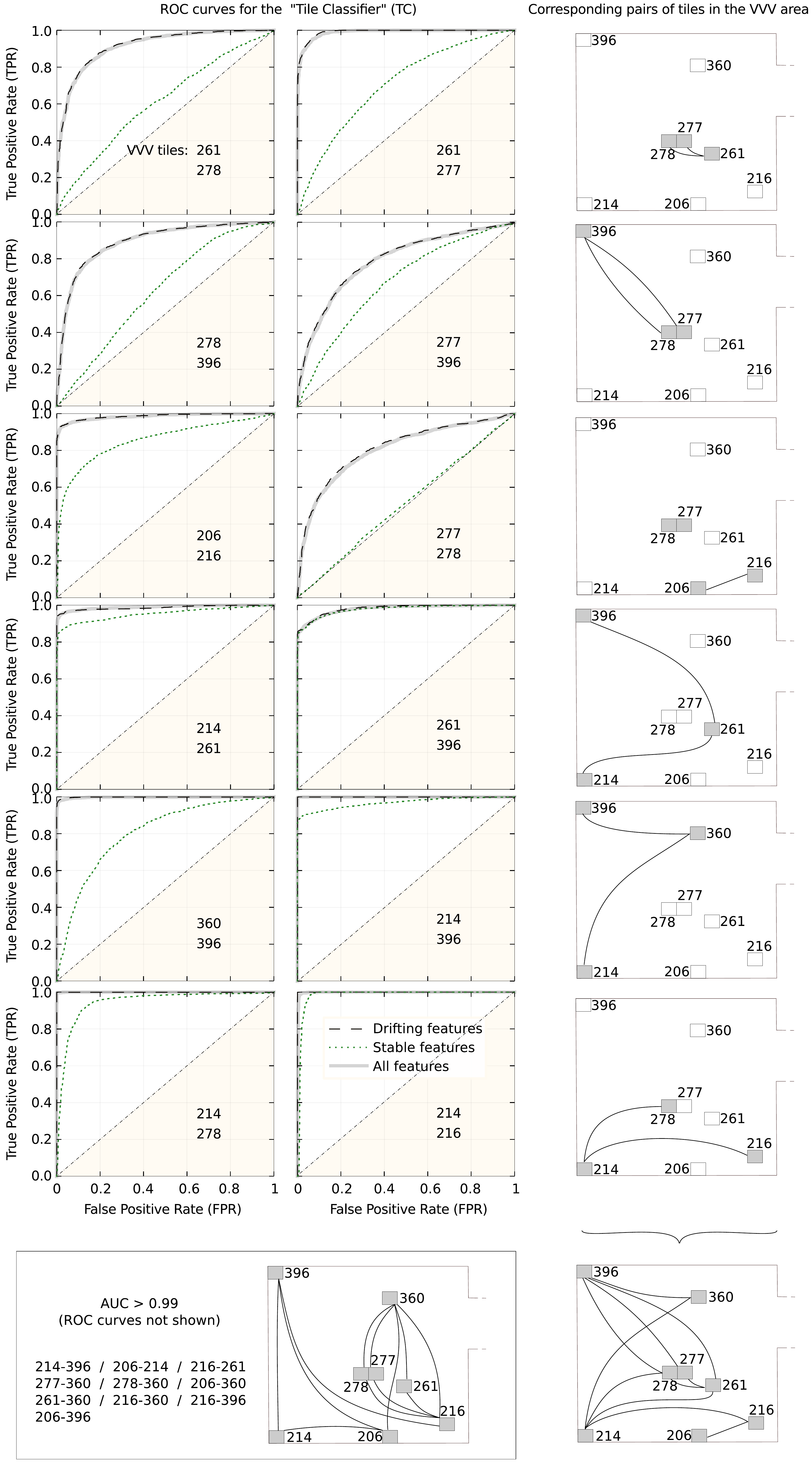}
    \caption{
        ROC curves for the combinations of two tiles selecting the more important features. Each panel has three curves: one using all the features (thick grey lines), the Drifting Features selected by RFE (dashed lines), and those considered stable, not selected by RFE (dotted lines).
    }
    \label{fig:nl_acurves_1}
\end{figure*}

We can arbitrarily divide the problem into ``few features" (4 or less selected features) and ``high number" (more than 4 selected features). Figure~\ref{fig:nl_acurves_1} shows ROC curves for the 12 ``high number" problems, and their relative locations in space. The figure shows curves for three classifiers: one trained with all the features in the dataset (All features), a second trained using only those selected by RFE, i.e. our candidates to Drifting Features, and a third one trained with those not selected by RFE (we call them the Stable subset). All the ``few features" subset cases (b216-b278 for example) produce trivial ROC curves saturated on the top-left corner for the three subsets, with $AUC>0.99$, which we do not show. 

Analyzing the results, the first observation is that, as expected, the classifier trained with the Drifting Features (those selected by feature selection) is always very similar in performance to the one trained with all the features. This result is a confirmation that RFECV does its work, selecting a subset of features that is responsible for the separation of the classes. The second result is that the performance of the models for the ``few features" problems is clearly superior to the ones shown in the figure, i.e. the ``high number" problems. This implies that the two behaviours noted in Figure  \ref{fig:lless_heatmap} correspond to problems that are easy to solve, where the separation is almost perfect and can be done with a few features, and problems that are harder, where the tiles cannot be fully separated and RFE selects bigger subsets. 

It is interesting to note the different response of the classifiers trained on the Stable subset on the easy and hard problems. In the hard problems RFECV selects a high number of features, which means that there are features with no considerable information about the tile of origin of the source. After several features were selected by RFECV, the rest of the features, the Stable subset, contains much less information about the origin and produces a classifier with low performance. For the easy problems, on the other hand, there are several features with much information about the tile of origin. Using just 2 to 4 features selected by RFECV is enough to produce an almost perfect classifier. The Stable subset in this case contains plenty of features with good information about the origin, producing a classifier with also almost perfect performance. If we take into account the relative positions of each pair of tiles for the Hard and Easy problems, there is not a definite pattern emerging. Most problems involving neighbour tiles, as for example b277, b278 and b261 are hard, and most problems involving tiles in the low-right region are easy.

\begin{figure*}
    \centering
    \includegraphics[width=.8\textwidth]{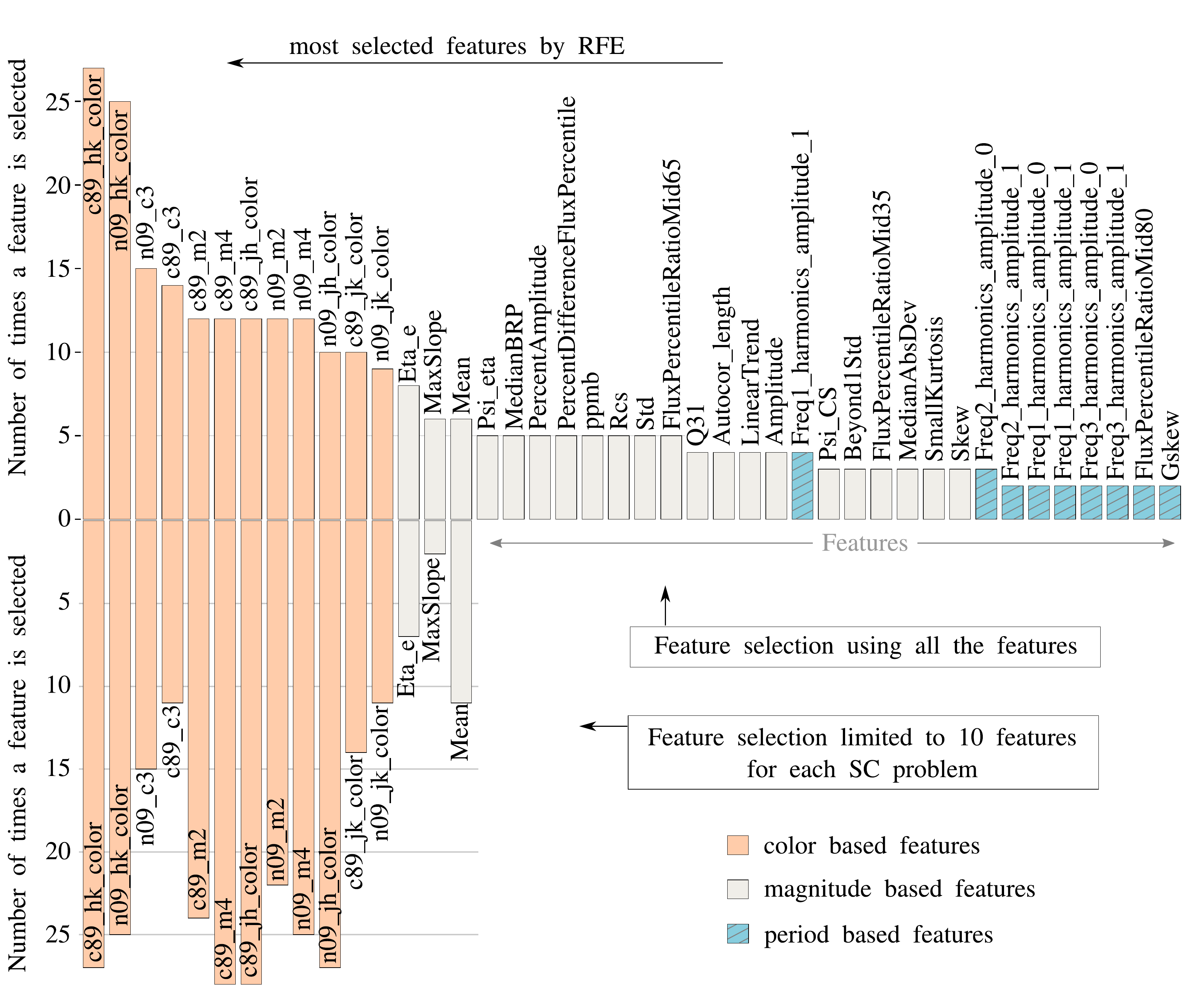}
    \caption{
        Total times that each feature was selected by RFECV over the 28 SC problems considered in this work. The different colors identify which of the three groups each feature belongs to: orange for those based on color, blue for those based on period, and white for those based only on magnitudes.
    }
    \label{fig:lless_freq}
\end{figure*}

Another relevant analysis for a feature selection method is which are the features that are selected in each case. The upper half of the Figure~\ref{fig:lless_freq} shows the number of times that each feature was selected by RFECV over the 28 TC problems, for all features that were selected at least two times (Table \ref{tab:rfe_no_limit_alt} in Appendix \ref{appendix:rfe} shows the list of features selected on each problem). Two particular features (\texttt{c89\_hk\_color} and \texttt{c09\_hk\_color}) were selected in almost all cases, while all the features related to pseudo-color \citep{cabral2020automatic} were the most frequently selected, appearing in at least half of the cases. This information suggests that color-related information in general is the most important characteristic to distinguish tiles.

\begin{figure}
    \centering
    \includegraphics[width=.41\textwidth]{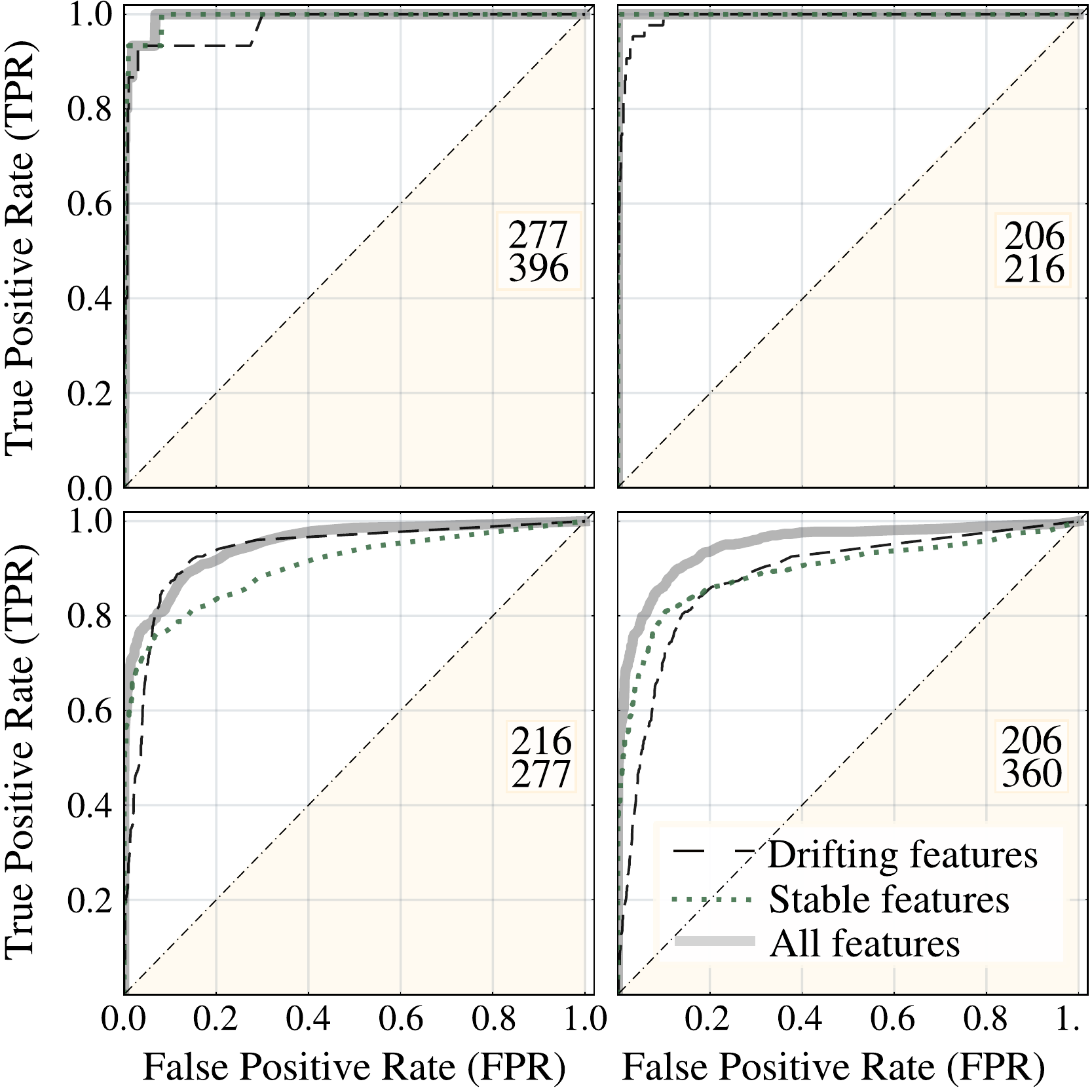}
    \caption{Same as Fig. \ref{fig:nl_acurves_1} for a fixed RFE selection of 10 features for the Drifting subset. Top row shows two Easy datasets while bottom row shows two Hard datasets.}
    \label{fig:l10_curve}
\end{figure}

The two very different behaviours of Hard and Easy problems will make harder the evaluation of the real influence of Drifting Features on the TC problems, because deleting only two features or half of the features will lead to diverse scenarios. To make an easier and fairer comparison we changed the feature selection method, using RFE with a fixed number of ten selected features.

The list of these selected features and their frequency can be observed in the bottom half of Figure~\ref{fig:lless_freq}.
Only 15 features were selected in total over the 28 TC problems, from whom the 11 more relevant are related to color, probably with a high dependence with the location of the tile.

The overall performance of the classifiers trained with the Full, Drifting and Stable subsets, for some exemplars of Easy and Hard datasets, can be seen in Figure~\ref{fig:l10_curve}. For the Easy problems (bottom row) we are now using less features in the Stable subset, leading to a lower performance. On the opposite side, for the hard problems (top row) we are now using more features in the Stable subset, leading to a clear increase in its performance. The rest of the TC problems show the same type of result (data not shown). 

Using a fixed selection of features with RFE we obtain in all cases a subset of 10 features (the Drifting subset) that can discriminate the tile of origin with high accuracy, and another subset (Stable) with much less information about the tile of origin of the sources. We analyze in the following the impact of these subsets on the SC problems.

\section{Evaluation of the influence of Drifting Features}
\label{section:evaluation}

In this section we evaluate the influence of the Drifting subsets selected in the previous step on the SC problems, i.e., to discriminate unknown sources and RRL variable stars.

For each pair of tiles we have three datasets, one with all the features (Full), a second with only 10 Drifting Features selected by RFE and finally one with the remaining features, the Stable subset. Different from the previous problem we have now, for the SC problem, two possibilities for each pair of tiles: first we train our classifiers in one of the tiles and search for RRLs in the other, and second we invert the tiles, training classifiers in the second tile of the pair and looking for RRLs in the first tile. Thus, for each one of the 56 SC problems we train corresponding RF classifiers and obtain three PR curves for the Full, Drifting and Stable classifiers. 

\begin{figure*}
    \centering
    \includegraphics[width=.95\textwidth]{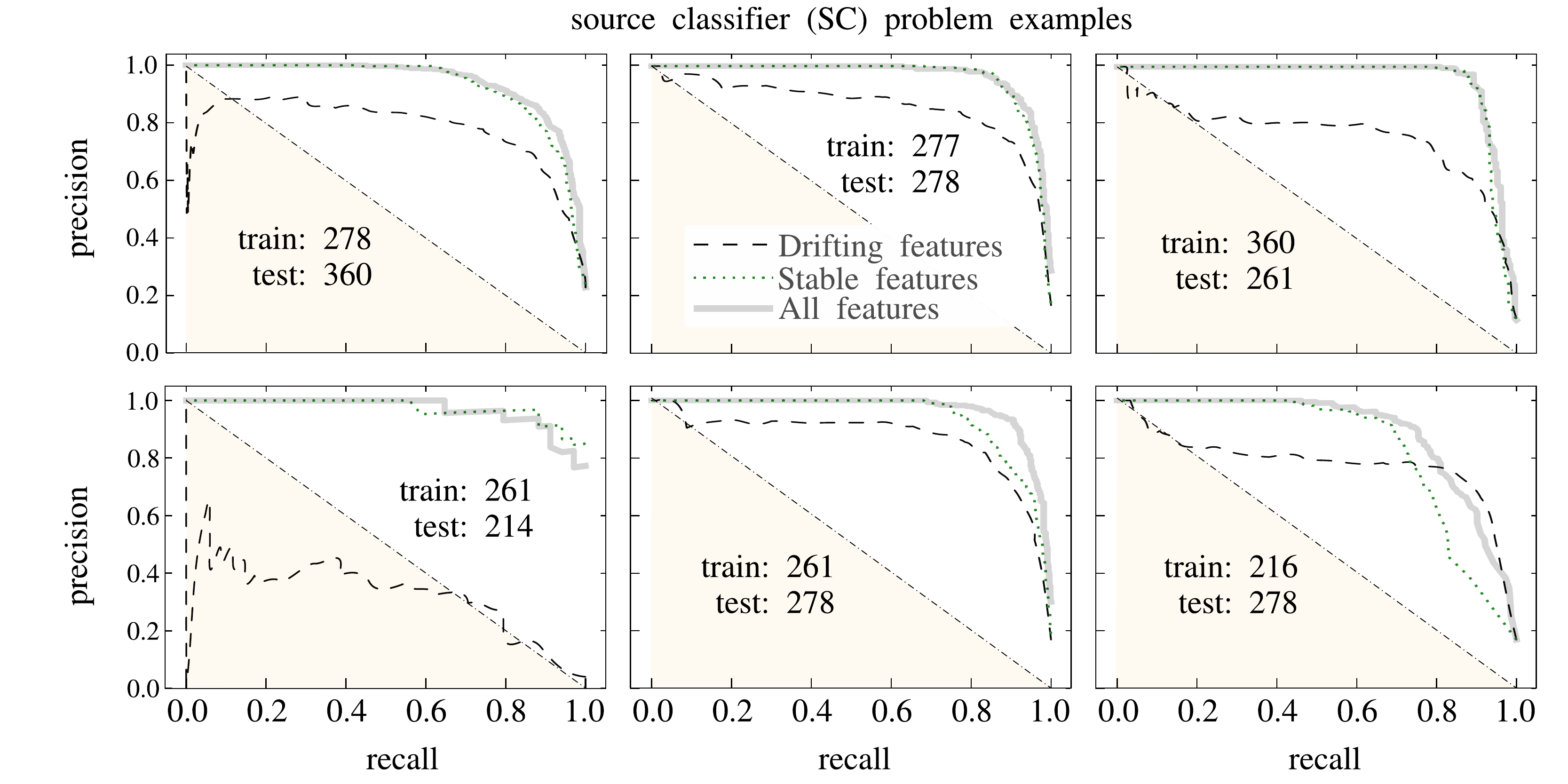}
    \caption{Precision-Recall curves for the SC problems. We show results for six combinations of train-test tiles
    using three classifiers trained with the Full, Drifting and Stable subsets of features.}
    \label{fig:nl_acurves_1_PR}
\end{figure*}

The complete results are presented in Appendix~\ref{appendix:evaluation}, while a summary of some representative cases can be seen in Figure~\ref{fig:nl_acurves_1_PR}. A first result is that, clearly, the Drifting subset shows lower performance in all cases. This was expected as most Drifting Features are related to color and \citet{cabral2020automatic} demonstrated that color alone cannot clearly identify RRLs. More interesting, if we compare the performance of the Full datasets with the Stable datasets, we can see that there is no clear advantage in eliminating the Drifting Features from the datasets. Full and Stable curves are almost similar in all cases. Differences are small and there is no clear pattern of when eliminating Drifting Features will improve the performance of the ML methods. 

\section{Discussion}
\label{section:conclusions}
In this work we introduced and discussed the concept of Drifting Features, related with small changes in the properties measured by the features, which can potentially harm the result of ML methods in astronomy. Using the identification of RRLs in VVV as a working example, we introduced a method for detecting Drifting Features, using an indirect ML method. We forced a classifier to learn the tile of origin of diverse sources, and select the features more relevant to this task as candidates to Drifting Features. We showed that this method can efficiently identify a reduced set of features that contains useful information about the tile of origin of the sources. We also showed that, for our particular example of detecting RRLs in VVV, Drifting Features are mostly related to color. On the other hand, we showed in the Section~\ref{section:evaluation} that, even if we have a clear set of Drifting Features in our problem, they are almost harmless for the identification of RRLs.

As a future work we will explore the influence of Drifting Features on the detection of other types of variable sources and other large scale ML experiments. We will also explore a different way of setting the number of selected features by RFE, considering all features that are relevant to the problem and not only the subset that shows the best performance for some metric or a fixed-length subset. 

% =============================================================================
% SECTION Acknowledgments
% =============================================================================

\begin{acknowledgements}
The authors would like to thank to their families and friends, and also IATE astronomers for useful comments and suggestions.
This work was partially supported by the Consejo Nacional
de Investigaciones Cient\'ificas y T\'ecnicas (CONICET, Argentina) and the Secretar\'ia de Ciencia y Tecnolog\'ia de la Universidad Nacional de C\'ordoba (SeCyT-UNC, Argentina).
J.B.C, are supported by a fellowship from CONICET.
Some processing was achieved with Argentine VO (NOVA) infrastructure, for which the authors express their gratitude.
We gratefully acknowledge data from the ESO Public Survey program ID 179.B-2002 taken with the VISTA telescope and products from the Cambridge Astronomical Survey Unit (CASU).
J.B.C. thanks to Maren Hempel by creating the template for the creation of the template on which the figure~\ref{fig:used_tiles} is based, and finally Bruno Sánchez and Martín Beroiz for the continuous support and friendship.
This research has made use of the
\url{http://adsabs.harvard.edu/}, Cornell University xxx.arxiv.org repository, adstex (\url{https://github.com/yymao/adstex}), astropy and the Python programming language.
\end{acknowledgements}


%\label{biblio}
\bibliographystyle{aa}
\bibliography{main.bib}
%%%%%%%%%%%%%%%%%%%%%%%%%%%%%%%%%%%%%%%%%%%%%%%%%%%%%%%%%%%%%%

% =============================================================================
% SECTION APPENDIXES
% =============================================================================

\appendix

\onecolumn
\section{Finding Drifting Features}
\label{appendix:rfe}

\begin{table}[tbh!]
\caption{RFE results without a minimum feature limit to try to identify the tile of a source. The rows contain the name of the features ordered from highest to lowest by the frequency they are selected, the columns are the datasets, and the body indicates with a check if that dataset selected that feature.}
    \centering
\resizebox{0.95\textwidth}{!}{
\small
   
\newcommand{\y}{$\checkmark$}
\newcommand{\n}{$\,\cdot$}
\newcommand{\colwd}{2pt}

\begin{tabular}{lp{\colwd}p{\colwd}p{\colwd}p{\colwd}p{\colwd}p{\colwd}p{\colwd}p{\colwd}p{\colwd}p{\colwd}p{\colwd}p{\colwd}p{\colwd}p{\colwd}p{\colwd}p{\colwd}p{\colwd}p{\colwd}p{\colwd}p{\colwd}p{\colwd}p{\colwd}p{\colwd}p{\colwd}p{\colwd}p{\colwd}p{\colwd}p{\colwd}r}
\toprule
%                                 & \multicolumn{3}{l}{b261} & \multicolumn{2}{l}{b277} & b278 & \multicolumn{7}{l}{b206} & \multicolumn{6}{l}{b216} & \multicolumn{5}{l}{b214} & b261 & b360 & b278 & \multicolumn{2}{l}{b277} \\
%                         Feature & b277 & b278 & b360 & b278 & b360 & b360 & b216 & b214 & b261 & b277 & b278 & b360 & b396 & b214 & b261 & b277 & b278 & b360 & b396 & b261 & b277 & b278 & b360 & b396 & b396 & b396 & b396 & b396 & Selected \\

%&\multicolumn{3}{l}{b261} & 
% \multicolumn{2}{l}{b277} & 
% b278 & 
% \multicolumn{7}{l}{b206} & 
% \multicolumn{6}{l}{b216} & 
% \multicolumn{5}{l}{b214} & 
% b261 & 
% b360 & 
% b278 & 
% \multicolumn{2}{l}{b277} \\

   Feature & \rotatebox{90}{b261 -- b277} & \rotatebox{90}{b261 -- b278} & \rotatebox{90}{b261 -- b360} & \rotatebox{90}{b277 -- b278} & \rotatebox{90}{b277 -- b360} & \rotatebox{90}{b278 -- b360} & \rotatebox{90}{b206 -- b216} & \rotatebox{90}{b206 -- b214} & \rotatebox{90}{b206 -- b261} & \rotatebox{90}{b206 -- b277} & \rotatebox{90}{b206 -- b278} & \rotatebox{90}{b206 -- b360} & \rotatebox{90}{b206 -- b396} & \rotatebox{90}{b216 -- b214} & \rotatebox{90}{b216 -- b261} & \rotatebox{90}{b216 -- b277} & \rotatebox{90}{b216 -- b278} & \rotatebox{90}{b216 -- b360} & \rotatebox{90}{b216 -- b396} &
   \rotatebox{90}{b214 -- b261} & \rotatebox{90}{b214 -- b277} &
   \rotatebox{90}{b214 -- b278} & \rotatebox{90}{b214 -- b360} & \rotatebox{90}{b214 -- b396} & \rotatebox{90}{b261 -- b396} & \rotatebox{90}{b360 -- b396} & \rotatebox{90}{b278 -- b396} & \rotatebox{90}{b277 -- b396} & \rotatebox{90}{Selected} \\

\midrule
%\noalign{\global\arrayrulewidth=0.1pt}
%\arrayrulecolor{tan}
                    c89\_hk\_color &   \y &  \y &    \y &    \y &    \y &    \y &    \y &    \y &    \y &    \y &    \y &    \y &    \y &    \y &    \y &    \y &    \y &    \y &    \y &    \y &    \y &    \y &    \y &    \y &    \y &    \y &    \y &    \n &       27 \\
                    n09\_hk\_color &    \y &    \y &    \y &    \y &    \y &    \n &    \y &    \y &    \y &    \y &    \y &    \y &    \y &    \y &    \y &    \y &    \y &    \y &    \y &    \y &    \y &    \y &    \n &    \y &    \n &    \y &    \y &    \y &       25 \\
                          n09\_c3 &    \y &    \y &    \y &    \y &    \y &    \y &    \y &    \n &    \n &    \n &    \n &    \n &    \n &    \n &    \n &    \n &    \n &    \n &    \n &    \y &    \n &    \y &    \y &    \y &    \y &    \y &    \y &    \y &       15 \\
                          c89\_c3 &    \y &    \y &    \y &    \y &    \y &    \y &    \y &    \n &    \n &    \n &    \n &    \n &    \n &    \n &    \n &    \n &    \n &    \n &    \n &    \y &    \n &    \y &    \y &    \y &    \n &    \y &    \y &    \y &       14 \\
                          c89\_m2 &    \y &    \y &    \n &    \y &    \y &    \n &    \y &    \n &    \n &    \n &    \n &    \n &    \n &    \n &    \n &    \n &    \n &    \n &    \n &    \y &    \n &    \y &    \n &    \y &    \y &    \y &    \y &    \y &       12 \\
                          c89\_m4 &    \y &    \y &    \n &    \y &    \y &    \n &    \y &    \n &    \n &    \n &    \n &    \n &    \n &    \n &    \n &    \n &    \n &    \n &    \n &    \y &    \n &    \y &    \n &    \y &    \y &    \y &    \y &    \y &       12 \\
                    c89\_jh\_color &    \y &    \y &    \n &    \y &    \n &    \n &    \y &    \n &    \n &    \n &    \n &    \n &    \n &    \y &    \n &    \n &    \n &    \n &    \n &    \y &    \n &    \y &    \n &    \y &    \y &    \y &    \y &    \y &       12 \\
                          n09\_m2 &    \y &    \y &    \n &    \y &    \y &    \n &    \y &    \n &    \n &    \n &    \n &    \n &    \n &    \n &    \n &    \n &    \n &    \n &    \n &    \y &    \n &    \y &    \n &    \y &    \y &    \y &    \y &    \y &       12 \\
                          n09\_m4 &    \y &    \y &    \n &    \y &    \y &    \n &    \y &    \n &    \n &    \n &    \n &    \n &    \n &    \n &    \n &    \n &    \n &    \n &    \n &    \y &    \n &    \y &    \n &    \y &    \y &    \y &    \y &    \y &       12 \\
                    n09\_jh\_color &    \y &    \y &    \n &    \y &    \n &    \n &    \y &    \n &    \n &    \n &    \n &    \n &    \n &    \y &    \n &    \n &    \n &    \n &    \n &    \y &    \n &    \y &    \n &    \y &    \n &    \y &    \y &    \n &       10 \\
                    c89\_jk\_color &    \y &    \y &    \n &    \y &    \n &    \n &    \y &    \n &    \n &    \n &    \n &    \n &    \n &    \n &    \n &    \n &    \n &    \n &    \n &    \y &    \n &    \y &    \n &    \y &    \n &    \y &    \y &    \y &       10 \\
                    n09\_jk\_color &    \y &    \y &    \n &    \y &    \n &    \n &    \y &    \n &    \n &    \n &    \n &    \n &    \n &    \n &    \n &    \n &    \n &    \n &    \n &    \y &    \n &    \y &    \n &    \n &    \n &    \y &    \y &    \y &        9 \\
                           Eta\_e &    \n &    \y &    \n &    \y &    \n &    \n &    \y &    \n &    \n &    \n &    \n &    \n &    \n &    \n &    \n &    \n &    \n &    \n &    \n &    \y &    \n &    \y &    \n &    \n &    \n &    \y &    \y &    \y &        8 \\
                        MaxSlope &    \n &    \n &    \n &    \n &    \n &    \n &    \y &    \n &    \n &    \n &    \n &    \n &    \n &    \n &    \n &    \n &    \n &    \n &    \n &    \n &    \n &    \y &    \n &    \y &    \n &    \y &    \y &    \y &        6 \\
                            Mean &    \n &    \y &    \n &    \y &    \n &    \n &    \y &    \n &    \n &    \n &    \n &    \n &    \n &    \n &    \n &    \n &    \n &    \n &    \n &    \n &    \n &    \n &    \n &    \n &    \n &    \y &    \y &    \y &        6 \\
                         Psi\_eta &    \n &    \y &    \n &    \n &    \n &    \n &    \y &    \n &    \n &    \n &    \n &    \n &    \n &    \n &    \n &    \n &    \n &    \n &    \n &    \n &    \n &    \n &    \n &    \n &    \n &    \y &    \y &    \y &        5 \\
                       MedianBRP &    \n &    \y &    \n &    \n &    \n &    \n &    \y &    \n &    \n &    \n &    \n &    \n &    \n &    \n &    \n &    \n &    \n &    \n &    \n &    \n &    \n &    \n &    \n &    \n &    \n &    \y &    \y &    \y &        5 \\
                PercentAmplitude &    \y &    \y &    \n &    \n &    \n &    \n &    \y &    \n &    \n &    \n &    \n &    \n &    \n &    \n &    \n &    \n &    \n &    \n &    \n &    \n &    \n &    \n &    \n &    \n &    \n &    \n &    \y &    \y &        5 \\
 PercentDifferenceFluxPercentile &    \n &    \y &    \n &    \y &    \n &    \n &    \y &    \n &    \n &    \n &    \n &    \n &    \n &    \n &    \n &    \n &    \n &    \n &    \n &    \n &    \n &    \n &    \n &    \n &    \n &    \n &    \y &    \y &        5 \\
                            ppmb &    \n &    \y &    \n &    \y &    \n &    \n &    \y &    \n &    \n &    \n &    \n &    \n &    \n &    \n &    \n &    \n &    \n &    \n &    \n &    \n &    \n &    \n &    \n &    \n &    \n &    \n &    \y &    \y &        5 \\
                             Rcs &    \y &    \y &    \n &    \y &    \n &    \n &    \n &    \n &    \n &    \n &    \n &    \n &    \n &    \n &    \n &    \n &    \n &    \n &    \n &    \n &    \n &    \n &    \n &    \n &    \n &    \n &    \y &    \y &        5 \\
                             Std &    \n &    \y &    \n &    \y &    \n &    \n &    \y &    \n &    \n &    \n &    \n &    \n &    \n &    \n &    \n &    \n &    \n &    \n &    \n &    \n &    \n &    \n &    \n &    \n &    \n &    \n &    \y &    \y &        5 \\
        FluxPercentileRatioMid65 &    \n &    \y &    \n &    \y &    \n &    \n &    \y &    \n &    \n &    \n &    \n &    \n &    \n &    \n &    \n &    \n &    \n &    \n &    \n &    \n &    \n &    \n &    \n &    \n &    \n &    \n &    \y &    \y &        5 \\
                             Q31 &    \n &    \y &    \n &    \n &    \n &    \n &    \y &    \n &    \n &    \n &    \n &    \n &    \n &    \n &    \n &    \n &    \n &    \n &    \n &    \n &    \n &    \n &    \n &    \n &    \n &    \n &    \y &    \y &        4 \\
                  Autocor\_length &    \n &    \y &    \n &    \n &    \n &    \n &    \y &    \n &    \n &    \n &    \n &    \n &    \n &    \n &    \n &    \n &    \n &    \n &    \n &    \n &    \n &    \n &    \n &    \n &    \n &    \n &    \y &    \y &        4 \\
                     LinearTrend &    \n &    \y &    \n &    \y &    \n &    \n &    \n &    \n &    \n &    \n &    \n &    \n &    \n &    \n &    \n &    \n &    \n &    \n &    \n &    \n &    \n &    \n &    \n &    \n &    \n &    \n &    \y &    \y &        4 \\
                       Amplitude &    \n &    \y &    \n &    \y &    \n &    \n &    \y &    \n &    \n &    \n &    \n &    \n &    \n &    \n &    \n &    \n &    \n &    \n &    \n &    \n &    \n &    \n &    \n &    \n &    \n &    \n &    \y &    \n &        4 \\
     Freq1\_harmonics\_amplitude\_1 &    \n &    \y &    \n &    \n &    \n &    \n &    \y &    \n &    \n &    \n &    \n &    \n &    \n &    \n &    \n &    \n &    \n &    \n &    \n &    \n &    \n &    \n &    \n &    \n &    \n &    \n &    \y &    \y &        4 \\
                          Psi\_CS &    \n &    \y &    \n &    \n &    \n &    \n &    \n &    \n &    \n &    \n &    \n &    \n &    \n &    \n &    \n &    \n &    \n &    \n &    \n &    \n &    \n &    \n &    \n &    \n &    \n &    \n &    \y &    \y &        3 \\
                      Beyond1Std &    \n &    \n &    \n &    \n &    \n &    \n &    \y &    \n &    \n &    \n &    \n &    \n &    \n &    \n &    \n &    \n &    \n &    \n &    \n &    \n &    \n &    \n &    \n &    \n &    \n &    \n &    \y &    \y &        3 \\
        FluxPercentileRatioMid35 &    \n &    \n &    \n &    \n &    \n &    \n &    \y &    \n &    \n &    \n &    \n &    \n &    \n &    \n &    \n &    \n &    \n &    \n &    \n &    \n &    \n &    \n &    \n &    \n &    \n &    \n &    \y &    \y &        3 \\
                    MedianAbsDev &    \n &    \n &    \n &    \n &    \n &    \n &    \y &    \n &    \n &    \n &    \n &    \n &    \n &    \n &    \n &    \n &    \n &    \n &    \n &    \n &    \n &    \n &    \n &    \n &    \n &    \n &    \y &    \y &        3 \\
                   SmallKurtosis &    \n &    \n &    \n &    \n &    \n &    \n &    \y &    \n &    \n &    \n &    \n &    \n &    \n &    \n &    \n &    \n &    \n &    \n &    \n &    \n &    \n &    \n &    \n &    \n &    \n &    \n &    \y &    \y &        3 \\
                            Skew &    \n &    \n &    \n &    \n &    \n &    \n &    \y &    \n &    \n &    \n &    \n &    \n &    \n &    \n &    \n &    \n &    \n &    \n &    \n &    \n &    \n &    \n &    \n &    \n &    \n &    \n &    \y &    \y &        3 \\
     Freq2\_harmonics\_amplitude\_0 &    \n &    \y &    \n &    \y &    \n &    \n &    \y &    \n &    \n &    \n &    \n &    \n &    \n &    \n &    \n &    \n &    \n &    \n &    \n &    \n &    \n &    \n &    \n &    \n &    \n &    \n &    \n &    \n &        3 \\
     Freq2\_harmonics\_amplitude\_1 &    \n &    \n &    \n &    \n &    \n &    \n &    \n &    \n &    \n &    \n &    \n &    \n &    \n &    \n &    \n &    \n &    \n &    \n &    \n &    \n &    \n &    \n &    \n &    \n &    \n &    \n &    \y &    \y &        2 \\
        FluxPercentileRatioMid80 &    \n &    \n &    \n &    \n &    \n &    \n &    \y &    \n &    \n &    \n &    \n &    \n &    \n &    \n &    \n &    \n &    \n &    \n &    \n &    \n &    \n &    \n &    \n &    \n &    \n &    \n &    \y &    \n &        2 \\
     Freq1\_harmonics\_amplitude\_2 &    \n &    \y &    \n &    \n &    \n &    \n &    \y &    \n &    \n &    \n &    \n &    \n &    \n &    \n &    \n &    \n &    \n &    \n &    \n &    \n &    \n &    \n &    \n &    \n &    \n &    \n &    \n &    \n &        2 \\
     Freq1\_harmonics\_rel\_phase\_1 &    \n &    \n &    \n &    \n &    \n &    \n &    \y &    \n &    \n &    \n &    \n &    \n &    \n &    \n &    \n &    \n &    \n &    \n &    \n &    \n &    \n &    \n &    \n &    \n &    \n &    \n &    \y &    \n &        2 \\
                           Gskew &    \n &    \n &    \n &    \n &    \n &    \n &    \n &    \n &    \n &    \n &    \n &    \n &    \n &    \n &    \n &    \n &    \n &    \n &    \n &    \n &    \n &    \n &    \n &    \n &    \n &    \n &    \y &    \y &        2 \\
     Freq3\_harmonics\_amplitude\_1 &    \n &    \y &    \n &    \n &    \n &    \n &    \y &    \n &    \n &    \n &    \n &    \n &    \n &    \n &    \n &    \n &    \n &    \n &    \n &    \n &    \n &    \n &    \n &    \n &    \n &    \n &    \n &    \n &        2 \\
     Freq3\_harmonics\_amplitude\_0 &    \n &    \n &    \n &    \n &    \n &    \n &    \y &    \n &    \n &    \n &    \n &    \n &    \n &    \n &    \n &    \n &    \n &    \n &    \n &    \n &    \n &    \n &    \n &    \n &    \n &    \n &    \n &    \y &        2 \\
                        PeriodLS &    \n &    \n &    \n &    \n &    \n &    \n &    \y &    \n &    \n &    \n &    \n &    \n &    \n &    \n &    \n &    \n &    \n &    \n &    \n &    \n &    \n &    \n &    \n &    \n &    \n &    \n &    \n &    \n &        1 \\
     Freq1\_harmonics\_rel\_phase\_2 &    \n &    \n &    \n &    \n &    \n &    \n &    \y &    \n &    \n &    \n &    \n &    \n &    \n &    \n &    \n &    \n &    \n &    \n &    \n &    \n &    \n &    \n &    \n &    \n &    \n &    \n &    \n &    \n &        1 \\
        FluxPercentileRatioMid50 &    \n &    \y &    \n &    \n &    \n &    \n &    \n &    \n &    \n &    \n &    \n &    \n &    \n &    \n &    \n &    \n &    \n &    \n &    \n &    \n &    \n &    \n &    \n &    \n &    \n &    \n &    \n &    \n &        1 \\
     Freq3\_harmonics\_amplitude\_2 &    \n &    \n &    \n &    \n &    \n &    \n &    \n &    \n &    \n &    \n &    \n &    \n &    \n &    \n &    \n &    \n &    \n &    \n &    \n &    \n &    \n &    \n &    \n &    \n &    \n &    \n &    \y &    \n &        1 \\
     Freq1\_harmonics\_amplitude\_0 &    \n &    \n &    \n &    \n &    \n &    \n &    \y &    \n &    \n &    \n &    \n &    \n &    \n &    \n &    \n &    \n &    \n &    \n &    \n &    \n &    \n &    \n &    \n &    \n &    \n &    \n &    \n &    \n &        1 \\
                      Period\_fit &    \n &    \n &    \n &    \n &    \n &    \n &    \y &    \n &    \n &    \n &    \n &    \n &    \n &    \n &    \n &    \n &    \n &    \n &    \n &    \n &    \n &    \n &    \n &    \n &    \n &    \n &    \n &    \n &        1 \\
     Freq2\_harmonics\_rel\_phase\_3 &    \n &    \n &    \n &    \n &    \n &    \n &    \n &    \n &    \n &    \n &    \n &    \n &    \n &    \n &    \n &    \n &    \n &    \n &    \n &    \n &    \n &    \n &    \n &    \n &    \n &    \n &    \y &    \n &        1 \\
     Freq3\_harmonics\_amplitude\_3 &    \n &    \n &    \n &    \n &    \n &    \n &    \y &    \n &    \n &    \n &    \n &    \n &    \n &    \n &    \n &    \n &    \n &    \n &    \n &    \n &    \n &    \n &    \n &    \n &    \n &    \n &    \n &    \n &        1 \\
     Freq1\_harmonics\_rel\_phase\_3 &    \n &    \n &    \n &    \n &    \n &    \n &    \n &    \n &    \n &    \n &    \n &    \n &    \n &    \n &    \n &    \n &    \n &    \n &    \n &    \n &    \n &    \n &    \n &    \n &    \n &    \n &    \y &    \n &        1 \\
     Freq3\_harmonics\_rel\_phase\_1 &    \n &    \n &    \n &    \n &    \n &    \n &    \y &    \n &    \n &    \n &    \n &    \n &    \n &    \n &    \n &    \n &    \n &    \n &    \n &    \n &    \n &    \n &    \n &    \n &    \n &    \n &    \n &    \n &        1 \\
     Freq2\_harmonics\_amplitude\_3 &    \n &    \n &    \n &    \n &    \n &    \n &    \y &    \n &    \n &    \n &    \n &    \n &    \n &    \n &    \n &    \n &    \n &    \n &    \n &    \n &    \n &    \n &    \n &    \n &    \n &    \n &    \n &    \n &        1 \\
     Freq2\_harmonics\_rel\_phase\_2 &    \n &    \n &    \n &    \n &    \n &    \n &    \n &    \n &    \n &    \n &    \n &    \n &    \n &    \n &    \n &    \n &    \n &    \n &    \n &    \n &    \n &    \n &    \n &    \n &    \n &    \n &    \n &    \n &        0 \\
                  PairSlopeTrend &    \n &    \n &    \n &    \n &    \n &    \n &    \n &    \n &    \n &    \n &    \n &    \n &    \n &    \n &    \n &    \n &    \n &    \n &    \n &    \n &    \n &    \n &    \n &    \n &    \n &    \n &    \n &    \n &        0 \\
     Freq2\_harmonics\_rel\_phase\_1 &    \n &    \n &    \n &    \n &    \n &    \n &    \n &    \n &    \n &    \n &    \n &    \n &    \n &    \n &    \n &    \n &    \n &    \n &    \n &    \n &    \n &    \n &    \n &    \n &    \n &    \n &    \n &    \n &        0 \\
     Freq2\_harmonics\_amplitude\_2 &    \n &    \n &    \n &    \n &    \n &    \n &    \n &    \n &    \n &    \n &    \n &    \n &    \n &    \n &    \n &    \n &    \n &    \n &    \n &    \n &    \n &    \n &    \n &    \n &    \n &    \n &    \n &    \n &        0 \\
     Freq3\_harmonics\_rel\_phase\_3 &    \n &    \n &    \n &    \n &    \n &    \n &    \n &    \n &    \n &    \n &    \n &    \n &    \n &    \n &    \n &    \n &    \n &    \n &    \n &    \n &    \n &    \n &    \n &    \n &    \n &    \n &    \n &    \n &        0 \\
        FluxPercentileRatioMid20 &    \n &    \n &    \n &    \n &    \n &    \n &    \n &    \n &    \n &    \n &    \n &    \n &    \n &    \n &    \n &    \n &    \n &    \n &    \n &    \n &    \n &    \n &    \n &    \n &    \n &    \n &    \n &    \n &        0 \\
                             Con &    \n &    \n &    \n &    \n &    \n &    \n &    \n &    \n &    \n &    \n &    \n &    \n &    \n &    \n &    \n &    \n &    \n &    \n &    \n &    \n &    \n &    \n &    \n &    \n &    \n &    \n &    \n &    \n &        0 \\
     Freq3\_harmonics\_rel\_phase\_2 &    \n &    \n &    \n &    \n &    \n &    \n &    \n &    \n &    \n &    \n &    \n &    \n &    \n &    \n &    \n &    \n &    \n &    \n &    \n &    \n &    \n &    \n &    \n &    \n &    \n &    \n &    \n &    \n &        0 \\
     Freq1\_harmonics\_amplitude\_3 &    \n &    \n &    \n &    \n &    \n &    \n &    \n &    \n &    \n &    \n &    \n &    \n &    \n &    \n &    \n &    \n &    \n &    \n &    \n &    \n &    \n &    \n &    \n &    \n &    \n &    \n &    \n &    \n &        0 \\
\bottomrule
\end{tabular}

}
\label{tab:rfe_no_limit_alt}
\end{table}

\begin{table}[tbh!]
\caption{RFE result selecting 10 most important features to try to identify a source tile. The rows contain the name of the features ordered from highest to lowest by the frequency they are selected, the columns the combination of dataset tiles and the body indicates with a check if that dataset selected that feature.}
    \centering
\resizebox{.95\textwidth}{!}{
\small
    \newcommand{\y}{$\checkmark$}
\newcommand{\n}{$\,\cdot$}
\newcommand{\colwd}{2pt}

\begin{tabular}{lp{\colwd}p{\colwd}p{\colwd}p{\colwd}p{\colwd}p{\colwd}p{\colwd}p{\colwd}p{\colwd}p{\colwd}p{\colwd}p{\colwd}p{\colwd}p{\colwd}p{\colwd}p{\colwd}p{\colwd}p{\colwd}p{\colwd}p{\colwd}p{\colwd}p{\colwd}p{\colwd}p{\colwd}p{\colwd}p{\colwd}p{\colwd}p{\colwd}r}
\toprule 

%b360 &b278&  &b277 & & & b261 &  &  &  & b214 &  &  &  &  & b216 &  &  &  &  &  & b206 &  &  &  &  &  &
%b396 &b396& b360 &b396 &b360 &b278 & b396 & b360 & b278 & b277 & b396 & b360 & b278 & b277 & b261 & b396 & b360 & b278 & b277 & b261 & b214 & b396 & b360 & b278 & b277 & b261 & b214 & b216

  Feature & \rotatebox{90}{b360 -- b396} & \rotatebox{90}{b278 -- b396}& \rotatebox{90}{b278 -- b360} &\rotatebox{90}{b277 -- b396} &\rotatebox{90}{b277 -- b360} &\rotatebox{90}{b277 -- b278} & \rotatebox{90}{b261 -- b396} & \rotatebox{90}{b261 -- b360} & \rotatebox{90}{b261 -- b278} & \rotatebox{90}{b261 -- b277} & \rotatebox{90}{b261 -- b396} & \rotatebox{90}{b261 -- b360} & \rotatebox{90}{b261 -- b278} & \rotatebox{90}{b261 -- b277} & \rotatebox{90}{b261 -- b261} & \rotatebox{90}{b216 -- b396} & \rotatebox{90}{b216 -- b360} & \rotatebox{90}{b216 -- b278} & \rotatebox{90}{b216 -- b277} & \rotatebox{90}{b216 -- b261} & \rotatebox{90}{b216 -- b214} & \rotatebox{90}{b206 -- b396} & \rotatebox{90}{b206 -- b360} & \rotatebox{90}{b206 -- b278} & \rotatebox{90}{b206 -- b277} & \rotatebox{90}{b206 -- b261} & \rotatebox{90}{b206 -- b214} & \rotatebox{90}{b206 -- b216}& \rotatebox{90}{Selected} \\

\midrule
c89\_m4                          &\y &\y &\y &\y &\y &\y &\y &\y &\y &\y &\y &\y &\y &\y &\y &\y &\y &\y &\y &\y &\y &\y &\y &\y &\y &\y &\y &\y &28 \\
c89\_jh\_color                   &\y &\y &\y &\y &\y &\y &\y &\y &\y &\y &\y &\y &\y &\y &\y &\y &\y &\y &\y &\y &\y &\y &\y &\y &\y &\y &\y &\y &28 \\
n09\_jh\_color                   &\y &\y &\y &\n &\y &\y &\y &\y &\y &\y &\y &\y &\y &\y &\y &\y &\y &\y &\y &\y &\y &\y &\y &\y &\y &\y &\y &\y &27 \\
c89\_hk\_color                   &\y &\y &\y &\n &\y &\y &\y &\y &\y &\y &\y &\y &\y &\y &\y &\y &\y &\y &\y &\y &\y &\y &\y &\y &\y &\y &\y &\y &27 \\
n09\_m4                          &\y &\y &\y &\y &\y &\y &\y &\y &\y &\y &\y &\y &\y &\y &\y &\y &\y &\y &\y &\y &\n &\y &\n &\y &\y &\y &\y &\n &25 \\
n09\_hk\_color                   &\y &\y &\y &\n &\y &\y &\n &\y &\n &\y &\y &\y &\y &\y &\y &\y &\y &\y &\y &\y &\y &\y &\y &\y &\y &\y &\y &\y &25 \\
c89\_m2                          &\y &\y &\y &\y &\y &\y &\y &\y &\y &\y &\y &\y &\y &\y &\n &\y &\y &\y &\y &\y &\y &\n &\y &\n &\n &\y &\y &\y &24 \\
n09\_m2                          &\y &\y &\y &\y &\y &\y &\y &\y &\y &\y &\y &\y &\y &\y &\y &\n &\n &\y &\y &\y &\y &\n &\n &\n &\n &\y &\y &\y &22 \\
n09\_c3                          &\y &\y &\y &\y &\y &\y &\y &\y &\y &\y &\y &\y &\y &\y &\y &\n &\n &\n &\n &\n &\n &\n &\n &\n &\n &\n &\n &\n &15 \\
c89\_jk\_color                   &\n &\y &\n &\n &\n &\n &\y &\n &\y &\n &\n &\n &\n &\n &\n &\y &\y &\y &\y &\y &\y &\y &\y &\y &\y &\n &\y &\n &14 \\
n09\_jk\_color                   &\n &\n &\n &\y &\n &\n &\y &\n &\y &\n &\n &\n &\n &\n &\n &\y &\y &\n &\y &\n &\y &\y &\y &\y &\y &\n &\n &\n &11 \\
c89\_c3                          &\y &\n &\y &\n &\y &\y &\n &\y &\n &\y &\y &\y &\y &\y &\y &\n &\n &\n &\n &\n &\n &\n &\n &\n &\n &\n &\n &\n &11 \\
Mean                             &\n &\n &\n &\y &\n &\n &\n &\n &\n &\n &\n &\n &\n &\n &\n &\y &\y &\y &\n &\y &\n &\y &\y &\y &\y &\y &\y &\n &11 \\
Eta\_e                           &\n &\n &\n &\n &\n &\n &\n &\n &\n &\n &\n &\n &\n &\n &\y &\n &\n &\n &\n &\n &\n &\y &\y &\y &\y &\y &\n &\y &7 \\
MaxSlope                         &\n &\n &\n &\n &\n &\n &\n &\n &\n &\n &\n &\n &\n &\n &\n &\n &\n &\n &\n &\n &\y &\n &\n &\n &\n &\n &\n &\y &2 \\
PercentDifferenceFluxPercentile  &\n &\n &\n &\y &\n &\n &\n &\n &\n &\n &\n &\n &\n &\n &\n &\n &\n &\n &\n &\n &\n &\n &\n &\n &\n &\n &\n &\n &1 \\
Std                              &\n &\n &\n &\y &\n &\n &\n &\n &\n &\n &\n &\n &\n &\n &\n &\n &\n &\n &\n &\n &\n &\n &\n &\n &\n &\n &\n &\n &1 \\
SmallKurtosis                    &\n &\n &\n &\n &\n &\n &\n &\n &\n &\n &\n &\n &\n &\n &\n &\n &\n &\n &\n &\n &\n &\n &\n &\n &\n &\n &\n &\y &1 \\
Skew                             &\n &\n &\n &\n &\n &\n &\n &\n &\n &\n &\n &\n &\n &\n &\n &\n &\n &\n &\n &\n &\n &\n &\n &\n &\n &\n &\n &\n &0 \\
Rcs                              &\n &\n &\n &\n &\n &\n &\n &\n &\n &\n &\n &\n &\n &\n &\n &\n &\n &\n &\n &\n &\n &\n &\n &\n &\n &\n &\n &\n &0 \\
Q31                              &\n &\n &\n &\n &\n &\n &\n &\n &\n &\n &\n &\n &\n &\n &\n &\n &\n &\n &\n &\n &\n &\n &\n &\n &\n &\n &\n &\n &0 \\
Psi\_eta                         &\n &\n &\n &\n &\n &\n &\n &\n &\n &\n &\n &\n &\n &\n &\n &\n &\n &\n &\n &\n &\n &\n &\n &\n &\n &\n &\n &\n &0 \\
Psi\_CS                          &\n &\n &\n &\n &\n &\n &\n &\n &\n &\n &\n &\n &\n &\n &\n &\n &\n &\n &\n &\n &\n &\n &\n &\n &\n &\n &\n &\n &0 \\
Period\_fit                      &\n &\n &\n &\n &\n &\n &\n &\n &\n &\n &\n &\n &\n &\n &\n &\n &\n &\n &\n &\n &\n &\n &\n &\n &\n &\n &\n &\n &0 \\
PeriodLS                         &\n &\n &\n &\n &\n &\n &\n &\n &\n &\n &\n &\n &\n &\n &\n &\n &\n &\n &\n &\n &\n &\n &\n &\n &\n &\n &\n &\n &0 \\
PercentAmplitude                 &\n &\n &\n &\n &\n &\n &\n &\n &\n &\n &\n &\n &\n &\n &\n &\n &\n &\n &\n &\n &\n &\n &\n &\n &\n &\n &\n &\n &0 \\
PairSlopeTrend                   &\n &\n &\n &\n &\n &\n &\n &\n &\n &\n &\n &\n &\n &\n &\n &\n &\n &\n &\n &\n &\n &\n &\n &\n &\n &\n &\n &\n &0 \\
MedianBRP                        &\n &\n &\n &\n &\n &\n &\n &\n &\n &\n &\n &\n &\n &\n &\n &\n &\n &\n &\n &\n &\n &\n &\n &\n &\n &\n &\n &\n &0 \\
MedianAbsDev                     &\n &\n &\n &\n &\n &\n &\n &\n &\n &\n &\n &\n &\n &\n &\n &\n &\n &\n &\n &\n &\n &\n &\n &\n &\n &\n &\n &\n &0 \\
LinearTrend                      &\n &\n &\n &\n &\n &\n &\n &\n &\n &\n &\n &\n &\n &\n &\n &\n &\n &\n &\n &\n &\n &\n &\n &\n &\n &\n &\n &\n &0 \\
Amplitude                        &\n &\n &\n &\n &\n &\n &\n &\n &\n &\n &\n &\n &\n &\n &\n &\n &\n &\n &\n &\n &\n &\n &\n &\n &\n &\n &\n &\n &0 \\
Gskew                            &\n &\n &\n &\n &\n &\n &\n &\n &\n &\n &\n &\n &\n &\n &\n &\n &\n &\n &\n &\n &\n &\n &\n &\n &\n &\n &\n &\n &0 \\
Autocor\_length                  &\n &\n &\n &\n &\n &\n &\n &\n &\n &\n &\n &\n &\n &\n &\n &\n &\n &\n &\n &\n &\n &\n &\n &\n &\n &\n &\n &\n &0 \\
Freq3\_harmonics\_rel\_phase\_3  &\n &\n &\n &\n &\n &\n &\n &\n &\n &\n &\n &\n &\n &\n &\n &\n &\n &\n &\n &\n &\n &\n &\n &\n &\n &\n &\n &\n &0 \\
Beyond1Std                       &\n &\n &\n &\n &\n &\n &\n &\n &\n &\n &\n &\n &\n &\n &\n &\n &\n &\n &\n &\n &\n &\n &\n &\n &\n &\n &\n &\n &0 \\
Con                              &\n &\n &\n &\n &\n &\n &\n &\n &\n &\n &\n &\n &\n &\n &\n &\n &\n &\n &\n &\n &\n &\n &\n &\n &\n &\n &\n &\n &0 \\
FluxPercentileRatioMid20         &\n &\n &\n &\n &\n &\n &\n &\n &\n &\n &\n &\n &\n &\n &\n &\n &\n &\n &\n &\n &\n &\n &\n &\n &\n &\n &\n &\n &0 \\
FluxPercentileRatioMid35         &\n &\n &\n &\n &\n &\n &\n &\n &\n &\n &\n &\n &\n &\n &\n &\n &\n &\n &\n &\n &\n &\n &\n &\n &\n &\n &\n &\n &0 \\
FluxPercentileRatioMid50         &\n &\n &\n &\n &\n &\n &\n &\n &\n &\n &\n &\n &\n &\n &\n &\n &\n &\n &\n &\n &\n &\n &\n &\n &\n &\n &\n &\n &0 \\
FluxPercentileRatioMid65         &\n &\n &\n &\n &\n &\n &\n &\n &\n &\n &\n &\n &\n &\n &\n &\n &\n &\n &\n &\n &\n &\n &\n &\n &\n &\n &\n &\n &0 \\
FluxPercentileRatioMid80         &\n &\n &\n &\n &\n &\n &\n &\n &\n &\n &\n &\n &\n &\n &\n &\n &\n &\n &\n &\n &\n &\n &\n &\n &\n &\n &\n &\n &0 \\
Freq1\_harmonics\_amplitude\_0   &\n &\n &\n &\n &\n &\n &\n &\n &\n &\n &\n &\n &\n &\n &\n &\n &\n &\n &\n &\n &\n &\n &\n &\n &\n &\n &\n &\n &0 \\
Freq1\_harmonics\_amplitude\_1   &\n &\n &\n &\n &\n &\n &\n &\n &\n &\n &\n &\n &\n &\n &\n &\n &\n &\n &\n &\n &\n &\n &\n &\n &\n &\n &\n &\n &0 \\
Freq1\_harmonics\_amplitude\_2   &\n &\n &\n &\n &\n &\n &\n &\n &\n &\n &\n &\n &\n &\n &\n &\n &\n &\n &\n &\n &\n &\n &\n &\n &\n &\n &\n &\n &0 \\
Freq1\_harmonics\_amplitude\_3   &\n &\n &\n &\n &\n &\n &\n &\n &\n &\n &\n &\n &\n &\n &\n &\n &\n &\n &\n &\n &\n &\n &\n &\n &\n &\n &\n &\n &0 \\
Freq1\_harmonics\_rel\_phase\_1  &\n &\n &\n &\n &\n &\n &\n &\n &\n &\n &\n &\n &\n &\n &\n &\n &\n &\n &\n &\n &\n &\n &\n &\n &\n &\n &\n &\n &0 \\
Freq1\_harmonics\_rel\_phase\_2  &\n &\n &\n &\n &\n &\n &\n &\n &\n &\n &\n &\n &\n &\n &\n &\n &\n &\n &\n &\n &\n &\n &\n &\n &\n &\n &\n &\n &0 \\
Freq1\_harmonics\_rel\_phase\_3  &\n &\n &\n &\n &\n &\n &\n &\n &\n &\n &\n &\n &\n &\n &\n &\n &\n &\n &\n &\n &\n &\n &\n &\n &\n &\n &\n &\n &0 \\
Freq2\_harmonics\_amplitude\_0   &\n &\n &\n &\n &\n &\n &\n &\n &\n &\n &\n &\n &\n &\n &\n &\n &\n &\n &\n &\n &\n &\n &\n &\n &\n &\n &\n &\n &0 \\
Freq2\_harmonics\_amplitude\_1   &\n &\n &\n &\n &\n &\n &\n &\n &\n &\n &\n &\n &\n &\n &\n &\n &\n &\n &\n &\n &\n &\n &\n &\n &\n &\n &\n &\n &0 \\
Freq2\_harmonics\_amplitude\_2   &\n &\n &\n &\n &\n &\n &\n &\n &\n &\n &\n &\n &\n &\n &\n &\n &\n &\n &\n &\n &\n &\n &\n &\n &\n &\n &\n &\n &0 \\
Freq2\_harmonics\_amplitude\_3   &\n &\n &\n &\n &\n &\n &\n &\n &\n &\n &\n &\n &\n &\n &\n &\n &\n &\n &\n &\n &\n &\n &\n &\n &\n &\n &\n &\n &0 \\
Freq2\_harmonics\_rel\_phase\_1  &\n &\n &\n &\n &\n &\n &\n &\n &\n &\n &\n &\n &\n &\n &\n &\n &\n &\n &\n &\n &\n &\n &\n &\n &\n &\n &\n &\n &0 \\
Freq2\_harmonics\_rel\_phase\_2  &\n &\n &\n &\n &\n &\n &\n &\n &\n &\n &\n &\n &\n &\n &\n &\n &\n &\n &\n &\n &\n &\n &\n &\n &\n &\n &\n &\n &0 \\
Freq2\_harmonics\_rel\_phase\_3  &\n &\n &\n &\n &\n &\n &\n &\n &\n &\n &\n &\n &\n &\n &\n &\n &\n &\n &\n &\n &\n &\n &\n &\n &\n &\n &\n &\n &0 \\
Freq3\_harmonics\_amplitude\_0   &\n &\n &\n &\n &\n &\n &\n &\n &\n &\n &\n &\n &\n &\n &\n &\n &\n &\n &\n &\n &\n &\n &\n &\n &\n &\n &\n &\n &0 \\
Freq3\_harmonics\_amplitude\_1   &\n &\n &\n &\n &\n &\n &\n &\n &\n &\n &\n &\n &\n &\n &\n &\n &\n &\n &\n &\n &\n &\n &\n &\n &\n &\n &\n &\n &0 \\
Freq3\_harmonics\_amplitude\_2   &\n &\n &\n &\n &\n &\n &\n &\n &\n &\n &\n &\n &\n &\n &\n &\n &\n &\n &\n &\n &\n &\n &\n &\n &\n &\n &\n &\n &0 \\
Freq3\_harmonics\_amplitude\_3   &\n &\n &\n &\n &\n &\n &\n &\n &\n &\n &\n &\n &\n &\n &\n &\n &\n &\n &\n &\n &\n &\n &\n &\n &\n &\n &\n &\n &0 \\
Freq3\_harmonics\_rel\_phase\_1  &\n &\n &\n &\n &\n &\n &\n &\n &\n &\n &\n &\n &\n &\n &\n &\n &\n &\n &\n &\n &\n &\n &\n &\n &\n &\n &\n &\n &0 \\
Freq3\_harmonics\_rel\_phase\_2  &\n &\n &\n &\n &\n &\n &\n &\n &\n &\n &\n &\n &\n &\n &\n &\n &\n &\n &\n &\n &\n &\n &\n &\n &\n &\n &\n &\n &0 \\
ppmb                             &\n &\n &\n &\n &\n &\n &\n &\n &\n &\n &\n &\n &\n &\n &\n &\n &\n &\n &\n &\n &\n &\n &\n &\n &\n &\n &\n &\n &0 \\

\bottomrule
\end{tabular}

}
\label{tab:ref_l10}
\end{table}

% ==========================================

\onecolumn
\section{Evaluation of the Drifting Features}
\label{appendix:evaluation}

\begin{figure}[tbh!]
    \centering
    \includegraphics[width=.97\textwidth]{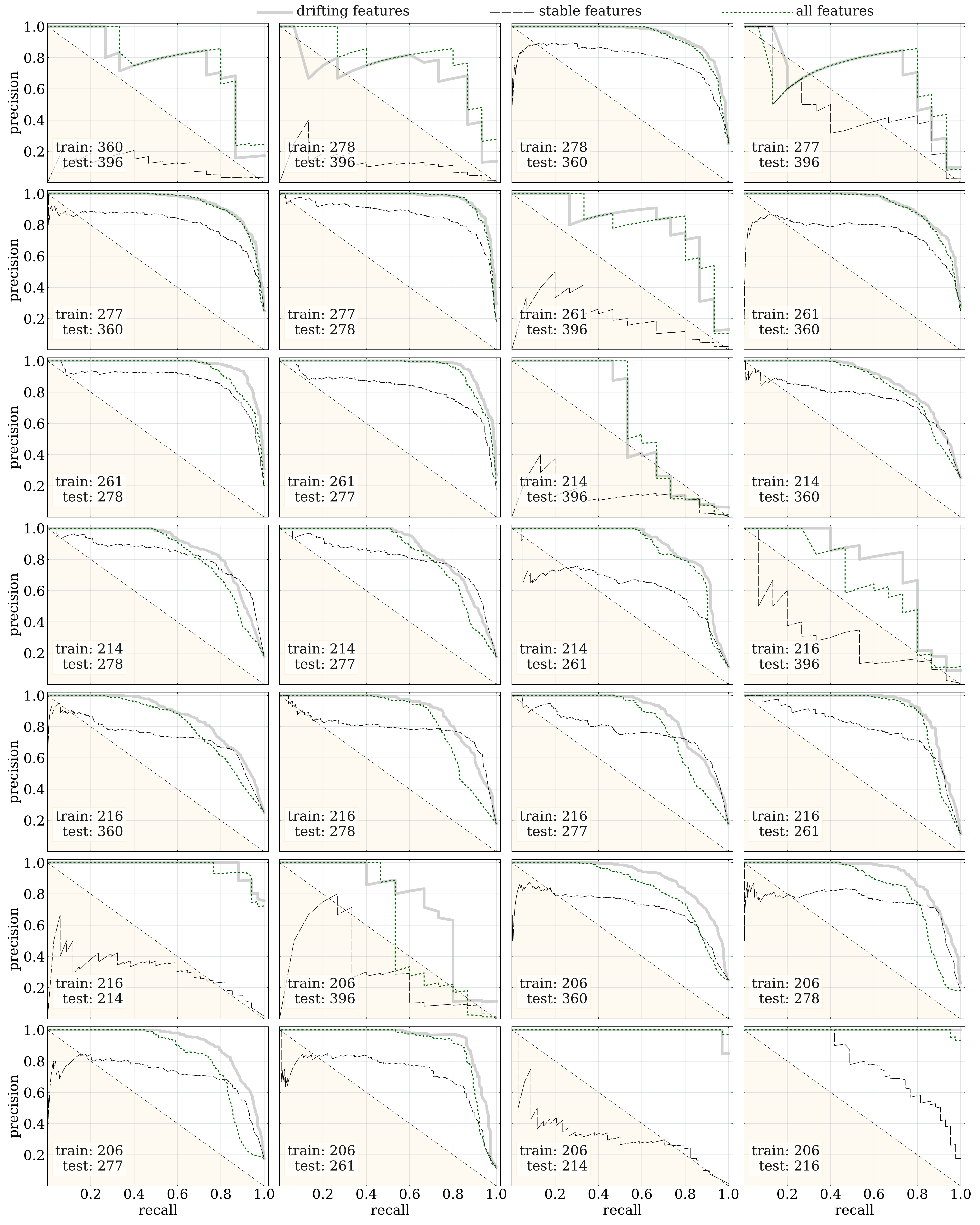}
    \caption{First 28 Precision-Recall curves for all combinations of train-test tiles for three different classifiers: one with all the features (Full), another only with the 10 Drifting Features found in that combination of train-test tiles (RFE), and finally one using the remaining using all but not the Drifting Features (No-RFE).}
    \label{fig:eva_acurves_1}
\end{figure}

\begin{figure}[tbh!]
    \centering
    \includegraphics[width=.97\textwidth]{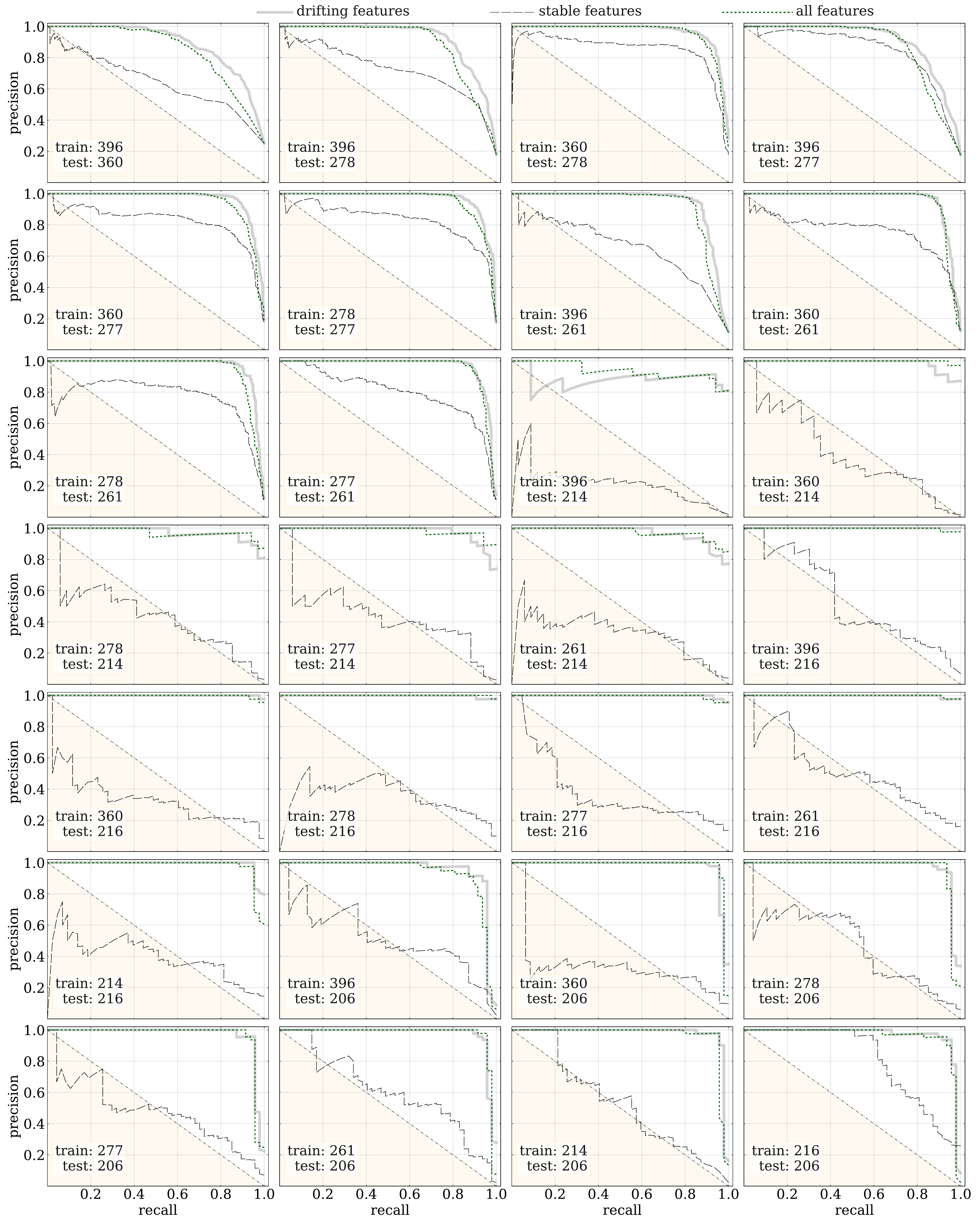}
    \caption{Last 28 Precision-Recall curves for all combinations of train-test tiles for three different classifiers: one with all the features (Full), another only with the 10 Drifting Features found in that combination of train-test tiles (RFE), and finally one using the remaining using all but not the Drifting Features (No-RFE).}
    \label{fig:eva_acurves_2}
\end{figure}

\end{document}